\documentclass[10pt,english,floatfix,nofootinbib,superscriptaddress,aps,prd,preprint, twocolumn]{revtex4}
\usepackage[utf8]{inputenc}
\usepackage{float}
\usepackage{array}
\usepackage{lipsum}
\usepackage{bbm}
\usepackage{dsfont}
\usepackage{graphicx}
\usepackage{amsmath}
\usepackage{tikz}
\usepackage{multirow}
\usepackage{braket}
\usepackage{bbm}
\usetikzlibrary{quotes,angles}
\usetikzlibrary{arrows} 
\usetikzlibrary{decorations.markings}
\usepackage{graphicx}
\usepackage[english]{babel}
\usepackage{color}
\usepackage{subfig}
\usepackage{caption}
\usepackage{tensor}
\usepackage{esint}
\usepackage[dvips]{epsfig}
\usepackage[dvips]{graphicx}
\usepackage{float}
\usepackage{units}
\usepackage{textcomp}
\usepackage{mathrsfs}
\usepackage{amsmath}
\usepackage[makeroom]{cancel}
\usepackage{amssymb}
\usepackage{amsbsy}
\usepackage{amsfonts}
\usepackage{amssymb,mathrsfs,xcolor}
\usepackage{esint}
\usepackage{array}
\usepackage{booktabs}
\usepackage{graphicx}
\usepackage{slashed}

\newcommand{\ie}{\begin{equation}}
\newcommand{\fe}{\end{equation}}
\newcommand{\se}{\begin{eqnarray}}
\newcommand{\ff}{\end{eqnarray}}

\usepackage{hyperref}
\hypersetup{colorlinks,breaklinks,
			citecolor=[rgb]{0,0.0,1.0},
            urlcolor=[rgb]{0.0,0.0,1.0},
            linkcolor=[rgb]{0,0.5,0.9}}

\begin{document}

\title{The role of non--metricity on neutrino behavior in bumblebee gravity}

\author{Yuxuan Shi}
\email{shiyx2280771974@gmail.com}
\affiliation{Department of Physics, East China University of Science and Technology, Shanghai 200237, China}

\author{A. A. Ara\'{u}jo Filho}
\email{dilto@fisica.ufc.br}
\affiliation{Departamento de Física, Universidade Federal da Paraíba, Caixa Postal 5008, 58051--970, João Pessoa, Paraíba,  Brazil.}


\date{\today}

\begin{abstract}

Within the context of bumblebee gravity, this work explores how \textit{non--metricity} alters the behavior and propagation of neutrinos. Our analysis is based on the black hole configuration introduced in Ref. \cite{filho2023vacuum}, focusing on how the spacetime deformation affects some neutrino--related processes. Three primary aspects are fundamentally taken into account: the modification in the energy deposition rate stemming from neutrino--antineutrino annihilation, the alterations in the oscillation phase caused by the background geometry, and the role of lensing effects on the transition probabilities among neutrino flavors. Complementing the analytical approach, numerical evaluations of oscillation probabilities are performed within a two--flavor scenario, accounting for both inverted and normal mass ordering configurations.

\end{abstract}

\maketitle


\section{Introduction}

Several gravitational and high--energy frameworks have pointed to the possibility that Lorentz symmetry might not hold in all regimes, particularly at extreme energy densities or in strong curvature backgrounds. Although experiments continue to confirm Lorentz invariance at accessible scales, this symmetry may no longer be preserved in more fundamental theories. In light of this, two main pathways through which Lorentz violation can emerge have been widely discussed \cite{bluhm2006overview}. On one hand, the explicit mechanism introduces symmetry--breaking terms directly into the action or equations of motion, leading to anisotropic effects that could, in principle, be measurable. On the other hand, spontaneous violation arises in a more subtle manner: while the underlying dynamics remain invariant, the vacuum configuration develops non--trivial structure that selects a preferred frame or direction \cite{colladay1998lorentz,bluhm2008spontaneous,kostelecky2004gravity}. A wide range of studies has tackled the consequences of both scenarios in modified gravity models and extended field theories \cite{3,5,4,7,1,8,ghosh2023does,2,6}.

The possibility that Lorentz symmetry may be spontaneously broken has motivated considerable exploration within alternative gravity models, especially in the context of the Standard Model Extension (SME) \cite{11,AraujoFilho:2024ykw,13,KhodadiPoDU2023,liu2024shadow,llvv2,Magalhaes:2025nql,filho2023vacuum,12,10,9,Liu:2024wpa,llvv1,Liu:2024oas,yang2023static,heidari2024scattering}. Among these approaches, the bumblebee framework stands out as a minimal and consistent setting where a vector field acquires a nonvanishing vacuum expectation value. As a result, local Lorentz symmetry is no longer preserved, and a preferred direction is spontaneously selected by the vacuum structure—not by explicit symmetry--breaking terms in the action \cite{bluhm2006overview}. Such models have proven effective in addressing various gravitational configurations, with particular attention given to how spontaneous Lorentz violation influences thermodynamic quantities and phase transitions in black hole solutions \cite{anacleto2018lorentz,araujo2022thermal,aa2021lorentz,araujo2021thermodynamic,araujo2021higher,reis2021thermal,araujo2022does,paperrainbow}.

Black hole configurations incorporating spontaneous Lorentz symmetry breaking—particularly within the scope of bumblebee gravity—have drawn increasing attention following the foundational solution introduced in Ref.~\cite{14}. Among the primary topics of interest is how such modified backgrounds influence radiative processes, including alterations in the spectrum of Hawking radiation due to deviations from standard Schwarzschild geometry \cite{kanzi2019gup}. The gravitational waves \cite{Khodadi:2025wuw} and gravitational baryogenesis \cite{Khodadi:2022mzt} were recently investigated.

The consequences for gravitational lensing have also been addressed, with particular focus on the angular deflection shifts resulting from the underlying anisotropy \cite{15}. Studies on the motion of matter around compact objects—especially the dynamics of accretion and radial infall—have further clarified the physical relevance of these deformations \cite{18,17}. In addition, the perturbative response of the spacetime has been investigated through the lens of quasinormal modes \cite{19,Liu:2022dcn} as well as Neutrino pair annihilation process \cite{Khodadi:2023yiw}. Recent developments have extended this analysis into the quantum regime, examining particle production processes near black hole horizons. These include scenarios involving both vector \cite{araujo2025does} and tensor fields \cite{AraujoFilho:2024ctw}.

The bumblebee gravity embedded in the \textit{metric--affine} formalism has opened new possibilities for examining how Lorentz--violating effects influence a range of gravitational and astrophysical phenomena \cite{filho2023vacuum}. Based on this latter reference, several works have applied this method to compute time delays in signal propagation, determine the spectrum of quasinormal modes, and evaluate angular deflections of light in deformed geometries \cite{gravitationaltraces}. The behavior of light in the strong deflection regime has also been analyzed in this context, revealing significant corrections to the standard lensing predictions \cite{araujo2024gravitational}. Efforts to generalize the static black hole solution to a rotating background have led to a Kerr--like extension within this symmetry-breaking scenario \cite{AraujoFilho:2024ykw}. Other lines of investigation have dealt with high-energy and observational features, such as quasiperiodic oscillations from galactic microquasars, modifications in the morphology of black hole shadows \cite{Gao:2024ejs}, and the internal structure of compact objects like strange quark stars and condensate dark stars \cite{Panotopoulos:2024jtn}. Further developments have tackled how these backgrounds affect scattering amplitudes \cite{heidari2024scattering}, the dynamics of accretion flows \cite{Lambiase:2023zeo}.

Furthermore, neutrinos display a set of quantum properties that sharply contrast with those of other fundamental particles, making them a subject of intense theoretical and experimental interest in particle physics \cite{neu42,neu43,neu44}. A defining feature of neutrino dynamics is the non--coincidence between their interaction basis—flavor states—and the basis in which they possess definite masses. Because of this misalignment, a neutrino created in a specific flavor state emerges as a linear combination of mass eigenstates. As it propagates, the phase evolution of each mass component differs, giving rise to an interference pattern that manifests as a periodic transformation among flavors \cite{neu40,neu39,neu41}. This quantum mechanical effect, known as neutrino oscillation, is a direct consequence of the correlation between mass splitting and quantum superposition.

Neutrino oscillations in flat spacetime are entirely dictated by the mass--squared differences between neutrino eigenstates, instead of their individual masses. These differences are defined through the relation
$$
\Delta m^2_{ij} = m_i^2 - m_j^2,  
$$
and the most frequently examined parameters include $|\Delta m^2_{21}|$, $|\Delta m^2_{31}|$, and $|\Delta m^2_{23}|$. Once the oscillation probabilities depend merely on these relative quantities, experimental observations are unable to determine the absolute neutrino mass scale. Rather, what is extracted from measurements are the gaps between squared masses, which are sufficient to describe the interference pattern underlying flavor transitions \cite{neu45}.

The presence of gravitational fields introduces nontrivial modifications to neutrino oscillation patterns, departing from the conventional flat--spacetime description, as one should expect. Unlike in Minkowski geometry, where flavor transitions depend exclusively on mass--squared differences, curved backgrounds contribute additional terms to the oscillation phase that can, in certain regimes, encode sensitivity to the absolute mass scale \cite{Shi:2025rfq,Shi:2025plr,Alloqulov:2024sns,Shi:2024flw,Chakrabarty:2023kld,AraujoFilho:2025rzh}. These gravitationally induced phase shifts become particularly relevant for ultra--relativistic neutrinos traveling across cosmological distances or escaping from compact astrophysical objects \cite{Shi:2023kid}. When neutrinos propagate through regions of non--Euclidean geometry, the cumulative phase they acquire reflects not only the properties of their mass eigenstates but also the curvature characteristics of the spacetime along their trajectory. Consequently, flavor transitions are modified in a way that standard flat--spacetime models fail to predict. Any deviation between the predicted and detected flavor distributions may thus indicate gravitationally induced modifications in the quantum evolution of neutrinos. This approach enables the extraction of information about the structure of the background spacetime while simultaneously providing an indirect way to access the absolute neutrino mass scale \cite{Shi:2023hbw,neu46,neu53,neu50,neu47,neu49,neu48,neu52,neu51}.

The evolution of neutrino flavors is deeply influenced by the curvature of the spacetime through which they move. Rather than being governed solely by intrinsic properties, their quantum phase is shaped by the geometry of the path, with the surrounding gravitational field leaving a direct fingerprint on their propagation \cite{neu54,Shi:2025rfq,neu55}. In strongly curved regions—particularly near dense and compact sources—gravitational lensing alters the trajectory of neutrinos, bending their paths. Such deviations affect the relative relation between mass eigenstates, modifying interference conditions and ultimately modifying the probability of flavor transitions \cite{neu53,Shi:2024flw,Shi:2025rfq,AraujoFilho:2025rzh}.

Recent research has emphasized how the underlying spacetime geometry affects neutrino oscillations, particularly through mechanisms such as gravitational lensing, which can interfere with the coherence between mass eigenstates essential for flavor transitions \cite{neu58,neu57,neu56}. In scenarios where the gravitational source is rotating, additional complications arise due to the influence of the source’s angular momentum on the neutrino phase. According to Swami’s findings, these rotational contributions depend explicitly on the spin of the massive object and can either suppress or amplify the oscillation probabilities, depending on the surrounding geometry. Such effects are especially pronounced near compact objects with masses on the order of that of the Sun \cite{neu59}, where the spacetime rotation significantly alters the neutrino’s quantum evolution.

A recent contribution in the literature explored how neutrino dynamics are affected by a black hole solution within bumblebee gravity formulated in the \textit{metric} formalism \cite{Shi:2025plr}. Motivated by this, the present work extends the investigation to the \textit{metric--affine} framework, where the effects of \textit{non--metricity} become relevant. In this context, we analyze how spontaneous Lorentz symmetry breaking—emerging from a black hole background in the bumblebee model \cite{filho2023vacuum}—modifies key aspects of neutrino behavior. Rather than limiting the discussion to propagation effects alone, the study focuses on three closely related processes: the energy deposition from neutrino–antineutrino annihilation, curvature--induced modifications to the oscillation phase, and the influence of gravitational lensing on flavor transition probabilities. The analysis is carried out using a two--flavor framework, incorporating both normal and inverted mass orderings in the numerical simulations. Furthermore, the results obtained here are systematically compared with those derived in the \textit{metric} formalism \cite{Shi:2025plr}, highlighting the distinct signatures introduced by \textit{non--metricity} in the \textit{metric--affine} scenario.

\section{The metric--affine bumblebee black hole}

As previously discussed, the focus of our investigation is a black hole solution derived in the context of bumblebee gravity constructed within the \textit{metric--affine} framework \cite{filho2023vacuum}
\begin{align}
\label{metricaffineeq}
\mathrm{d}s^{2} = & -\dfrac{1-\dfrac{2M}{r}}{\sqrt{X_1X_2}}\mathrm{d}t^{2} + \sqrt{\dfrac{X_1}{X_2^3}}\left(1-\dfrac{2M}{r}\right)^{-1}\mathrm{d}r^{2}\\
& + r^{2}\mathrm{d}\theta^{2} + r^{2} \sin^{2}\theta\,\mathrm{d}\varphi^{2}
\nonumber.
\end{align}
where
\begin{align}
X_1 &= 1 + \dfrac{3X}{4}\\
X_2 &= 1 - \dfrac{X}{4}
\end{align}

The forthcoming section is dedicated to examining the energy deposited through neutrino--antineutrino annihilation in the black hole’s vicinity. It then moves on to assess how the background geometry modifies the oscillation phase and affects the probability of flavor transitions. In sequence, the effects of gravitational deflection on neutrino trajectories are evaluated, emphasizing how curvature distorts their paths. To reinforce our theoretical perspectives, numerical results are presented as well.

Before proceeding, we clarify the scope of Lorentz symmetry breaking considered in this work. The present analysis is restricted to the gravitational sector of the SME as realized in the metric--affine bumblebee model. In this framework, Lorentz violation arises spontaneously through the vacuum expectation value of the bumblebee field and manifests itself exclusively through modifications of the background geometry, encoded in the parameter $X=\xi b^{2}$. Here, $\xi$ is a dimensionless coupling constant, while $b_{\mu}$ represents the vacuum expectation value of the bumblebee field responsible for spontaneous Lorentz symmetry breaking.
 
We do not include independent matter--sector SME coefficients (such as $u_{\mu}$, $s_{\mu\nu}$, or related operators) that would directly modify the fermionic kinetic terms and, consequently, the particle dispersion relations. Neutrinos are therefore assumed to obey the standard dispersion relation in curved spacetime, so that all deviations from the general relativistic case originate solely from the Lorentz--violating gravitational background. Including additional matter--sector Lorentz--violating effects would lead to further corrections to the propagation and oscillation phase, which lie beyond the scope of the present work.


\section{Energy deposition rate}

This section investigates the mechanism of energy deposition in a gravitational setting altered fundamentally by the Lorentz--violating coefficient $X$, introduced in Eq.~(\ref{metricaffineeq}). The dominant contribution to this process arises from the annihilation of neutrino--antineutrino pairs in the vicinity of the compact object. The energy release rate, evaluated per unit volume and time, is determined by the following expression \cite{Salmonson:1999es}:
\ie
\dfrac{\mathrm{d}\mathrm{E}(r)}{\mathrm{d}t\mathrm{d}V}=2 \, \Tilde{K} \,\mathrm{G}_{f}^{2}\, \mathrm{f}(r)\iint
\overset{\nsim}{n}(\varepsilon_{\nu})\overset{\nsim}{n}(\varepsilon_{\overline{\nu}})
(\varepsilon_{\nu} + \varepsilon_{\overline{\nu}})
\varepsilon_{\nu}^{3}\varepsilon_{\overline{\nu}}^{3}
\mathrm{d}\varepsilon_{\nu}\mathrm{d} \varepsilon_{\overline{\nu}}
\fe
in which we have
\ie
\Tilde{K} = \dfrac{1}{6\pi}(1\pm4\sin^{2}\theta_{W}+8\sin^{4} \theta_{W}).
\fe

Explicit expressions for the energy deposition rates arising from different annihilation (neutrino--antineutrino) channels can be obtained by setting the Weinberg angle to its standard value, $\sin^2\theta_W = 0.23$. In other words, these formulas reveal how both the weak interaction strength and the flavor configuration of the annihilating pair govern the efficiency of energy transfer in such processes
\ie
\Tilde{K}(\nu_{\mu},\overline{\nu}_{\mu}) = \Tilde{K}(\nu_{\tau},\overline{\nu}_{\tau})
=\dfrac{1}{6\pi}\left(1-4\sin^{2}\theta_{W} + 8\sin^{4}\theta_{W}\right),
\fe
and, analogously, we write
\ie
\Tilde{K}(\nu_{e},\overline{\nu}_{e})
=\dfrac{1}{6\pi}\left(1+4\sin^{2}\theta_{W} + 8\sin^{4}\theta_{W}\right).
\fe

The determination of energy deposition rates for the corresponding annihilation processes begins by fixing the parameters that govern weak interactions. In this analysis, the Fermi coupling constant is set to $\mathrm{G}_f^2 = 5.29 \times 10^{-44}\,\mathrm{cm^2MeV^{-2}}$, while the weak mixing angle is taken as $\sin^2\theta_W = 0.23$, which defines the relative strength of the interactions. Notice that these constants are essential for distinguishing the contributions from different flavor combinations. In this manner, by accomplishing the angular integration over the interaction cross sections, one arrives at an expression for the deposited energy that aligns with the structure reported in Ref.~\cite{Salmonson:1999es}
\begin{align}
\mathrm{f}(r)&=\iint\left(1-\bm{\overset{\nsim}{\Omega}_{\nu}}\cdot\bm{\overset{\nsim}{\Omega}_{\overline{\nu}}}\right)^{2}
\mathrm{d}\overset{\nsim}{\Omega}_{\nu}\mathrm{d}\overset{\nsim}{\Omega}_{\overline{\nu}}\notag\\
&=\dfrac{2\pi^{2}}{3}(1 - x)^{4}\left(x^{2} + 4x + 5\right)
\end{align}
where $x$ is defined as follows
\ie
x = \sin\theta_{r}.
\fe

The angle $\theta_r$ defines the deviation between a particle’s direction of motion and the tangential orientation of a circular orbit at a specific radial position $r$. The trajectories of neutrinos and antineutrinos are described by the normalized vectors $\overset{\nsim}{\Omega}_{\nu}$ and $\overset{\nsim}{\Omega}_{\overline{\nu}}$, respectively, while their directional integration is carried out over the solid angle elements $\mathrm{d}\overset{\nsim}{\Omega}_{\nu}$ and $\mathrm{d}\overset{\nsim}{\Omega}_{\overline{\nu}}$. Under thermal equilibrium at temperature $\mathrm{T}$, their phase--space occupation numbers, $\overset{\nsim}{n}(\varepsilon_{\nu})$ and $\overset{\nsim}{n}(\varepsilon_{\overline{\nu}})$, follow Fermi--Dirac distributions, consistent with the statistical treatment outlined in Ref.~\cite{Salmonson:1999es}
\ie
\overset{\nsim}{n}(\varepsilon_{\nu}) = \frac{2}{h^{3}}\dfrac{1}{e^{\left({\frac{\varepsilon_{\nu}}{k \, \mathrm{T}}}\right)} + 1}.
\fe

The corresponding evaluation of energy deposition is carried out through a formulation that incorporates the fundamental constants $h$ (Planck) and $k$ (Boltzmann), which encode the quantum and thermal aspects of the process. Based on this setup, one obtains an expression for the energy transfer rate, computed per unit volume and time. In this manner, we follow the procedure presented in Ref.~\cite{Salmonson:1999es}
\ie
\frac{\mathrm{d}\mathrm{E}}{\mathrm{d}t\mathrm{d}V} = \frac{21\zeta(5)\pi^{4}}{h^{6}}\Tilde{K} \, \mathrm{G}_{f}^{2} \, \mathrm{f}(r)(k \, \mathrm{T})^{9}.
\fe

The temperature recorded by an observer at a radial location $r$ is subject to gravitational redshift effects, which impose the condition $\mathrm{T}(r)\sqrt{-\mathrm{g}_{tt}(r)} = \text{const}$. This relation reflects the way in which spacetime curvature alters thermal measurements within a gravitational field \cite{Salmonson:1999es}. At the boundary of the neutrinosphere, specified by $r = \mathrm{R}$, the emission temperature for neutrinos is constrained by this redshift condition and must satisfy the corresponding expression \cite{Salmonson:1999es}
\ie
\mathrm{T}(r)\sqrt{-\mathrm{g}_{tt}(r)} = \mathrm{T}(\mathrm{R})\sqrt{-\mathrm{g}_{tt}(\mathrm{R})}.
\fe  
In this context, $\mathrm{R}$ designates the radius of the compact object generating the gravitational background. To facilitate the analysis, the temperature distribution $\mathrm{T}(r)$ is reformulated using the gravitational redshift condition previously discussed. By incorporating it directly into the temperature dependence, one obtains a revised expression for the neutrino luminosity that accounts for the influence of curvature on thermal emission \cite{Salmonson:1999es}
\ie
\mathrm{L}_{\infty} = -\mathrm{g}_{tt}(\mathrm{R})L(\mathrm{R}).
\fe

The luminosity for an individual neutrino flavor, assessed at the location of the neutrinosphere (i.e., at $r = \mathrm{R}$), can be expressed through the following relation, which incorporates the relevant thermal and geometric factors \cite{Salmonson:1999es}:
\ie
\mathrm{L}(\mathrm{R}) = 4 \pi \mathrm{R}_{0}^{2}\dfrac{7}{4}\dfrac{a\,c}{4}\mathrm{T}^{4}(\mathrm{R}).
\fe

The constants $c$ and $a$ appearing in the expression represent, respectively, the speed of light in vacuum and the radiation constant. To incorporate the influence of spacetime curvature, the temperature measured at a specific radius must be corrected for gravitational redshift. This correction establishes a proper correlation between the thermal profile and the underlying geometry of the background spacetime \cite{Salmonson:1999es}:
\ie
\begin{split}
\frac{\mathrm{d}\mathrm{E}(r)}{\mathrm{d}t \, \mathrm{d}V} & = \dfrac{21\zeta(5)\pi^{4}}{h^{6}}
\Tilde{K} \, \mathrm{G}_{f}^{2} \, k^{9}\left(\dfrac{7}{4}\pi a\,c\right)^{-\frac{9}{4}}\\
& \times \mathrm{L}_{\infty}^{\frac{9}{4}}\mathrm{f}(r)
\left[\dfrac{\sqrt{-\mathrm{g}_{tt}(\mathrm{R})}}{-\mathrm{g}_{tt}(r)}\right]^{\frac{9}{2}} \mathrm{R}^{-\frac{9}{2}}.
\end{split}
\fe

One should emphasize that the symbol $\zeta(s)$ in the expression above denotes the Riemann zeta function. When the argument satisfies $s > 1$, this function is represented by the following series (infinite):
\ie
\zeta(s) = \sum_{n=1}^{\infty} \frac{1}{n^s}.
\fe

The evaluation of energy deposition in curved spacetime requires more than just accounting for the radial dependence— as we mentioned early, it also demands careful consideration of the spacetime geometry itself—, particularly the behavior of the metric components near the compact object's boundary. To obtain the total radiated energy, one must integrate the local deposition rate over time. A key step in this procedure is the determination of the angular function $\mathrm{f}(r)$, which involves a detailed reformulation of the auxiliary variable $x$ introduced earlier during the angular integration. The full treatment proceeds through the following steps \cite{Salmonson:1999es,Shi:2023kid,AraujoFilho:2024mvz,Lambiase:2020iul}
\begin{align}
x^{2}& = \sin^{2}\theta_{r}|_{\theta_{\mathrm{R}}=0}\notag\\
&=1-\dfrac{\mathrm{R}^{2}}{r^{2}}\dfrac{\mathrm{g}_{tt}(r)}{\mathrm{g}_{tt}(\mathrm{R})}.
\end{align}

Evaluating the total energy deposited around a compact object involves integrating the energy density—defined per unit time and per unit volume—throughout the spatial region influenced by the gravitational field. This integration crucially depends on the angular factor, which encapsulates how spacetime curvature affects the trajectories and interactions of the involved particles. The behavior of this angular component is intrinsically governed by the specific structure of the underlying metric \cite{Shi:2023kid,Lambiase:2020iul,AraujoFilho:2025rzh,Shi:2025rfq}
\begin{align}
\dot{Q} & = \frac{\mathrm{d}\mathrm{E}}{\sqrt{-\mathrm{g}_{tt}(r)}\mathrm{d}t}\\
&=\dfrac{84\zeta(5)\pi^{5}}{h^{6}}\Tilde{K}\, \mathrm{G}_{f}^{2} \, k^{9}
\left(\dfrac{7}{4}\pi a\,c\right)^{-\frac{9}{4}}
\mathrm{L}_{\infty}^{\frac{9}{4}}\left[-\mathrm{g}_{tt}(\mathrm{R})\right]^{\frac{9}{4}}\\
& \times 
\mathrm{R}^{-\frac{3}{2}}\int_{1}^{\infty}(x-1)^4\left(x^2+4x+5\right)\sqrt{\dfrac{\mathrm{g}_{rr}(y \mathrm{R})}{-\mathrm{g}_{tt}^9(y \mathrm{R})}}y^2\mathrm{d}y.
\end{align}

The total energy deposition rate, denoted by $\dot{Q}$, quantifies how much neutrino energy is converted into electron--positron pairs at a particular radial location \cite{Salmonson:1999es}. When this rate becomes sufficiently high, it can give rise to relevant astrophysical phenomena, driven by intense pair production. To better understand the impact of gravity on this mechanism, it is essential to contrast the relativistic and Newtonian treatments of energy deposition. This comparison makes it possible to isolate the influence of curvature and reveals how general relativistic corrections affect the efficiency of neutrino--mediated energy transfer \cite{Salmonson:1999es,Lambiase:2020iul,Shi:2023kid,shi2022neutrino}
\ie
\begin{split}
\label{ratio_Q}
 \frac{\dot{Q}}{\dot{Q}_{\text{Newton}}} = 3\left[-\mathrm{g}_{tt}(\mathrm{R})\right]^{\frac{9}{4}}
& \int_{1}^{\infty}(x - 1)^{4}\left(x^{2} + 4x + 5\right) \\
& \times \sqrt{\dfrac{\mathrm{g}_{rr}(y \mathrm{R})}{-\mathrm{g}_{tt}(y \mathrm{R})}}y^2\mathrm{d}y.
\end{split}
\fe
where the follow definitions are shown below
\ie
\begin{split}
\mathrm{g}_{tt}(\mathrm{R})&= -\dfrac{1-\dfrac{2M}{\mathrm{R}}}{\sqrt{X_1X_2}},\\
\mathrm{g}_{tt}(y \mathrm{R})&= -\dfrac{1-\dfrac{2M}{y \mathrm{R}}}{\sqrt{X_1X_2}}.
\end{split}
\fe
In addition to them, we have
\begin{align}
x^{2}=1-\dfrac{1}{y^{2}}\frac{1-\dfrac{2M}{y \mathrm{R}}}{1-\dfrac{2M}{\mathrm{R}}}.
\end{align}

\begin{figure*}
\centering
\includegraphics[height=6.5cm]{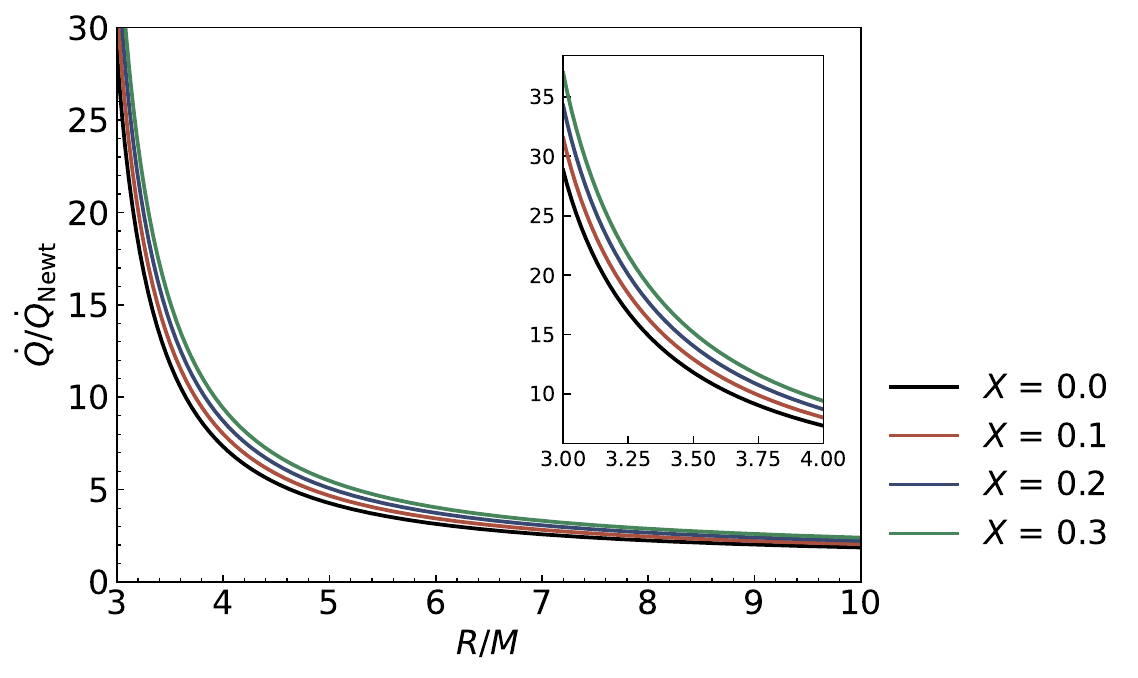}
\caption{The graph illustrates how the ratio $\dot{Q}/\dot{Q}_{\text{Newton}}$ varies with respect to the quantity $R/M$, for several chosen values of the Lorentz--violating parameter $X$.}
\label{depprate}
\end{figure*}

\begin{figure*}
\centering
\includegraphics[height=6.5cm]{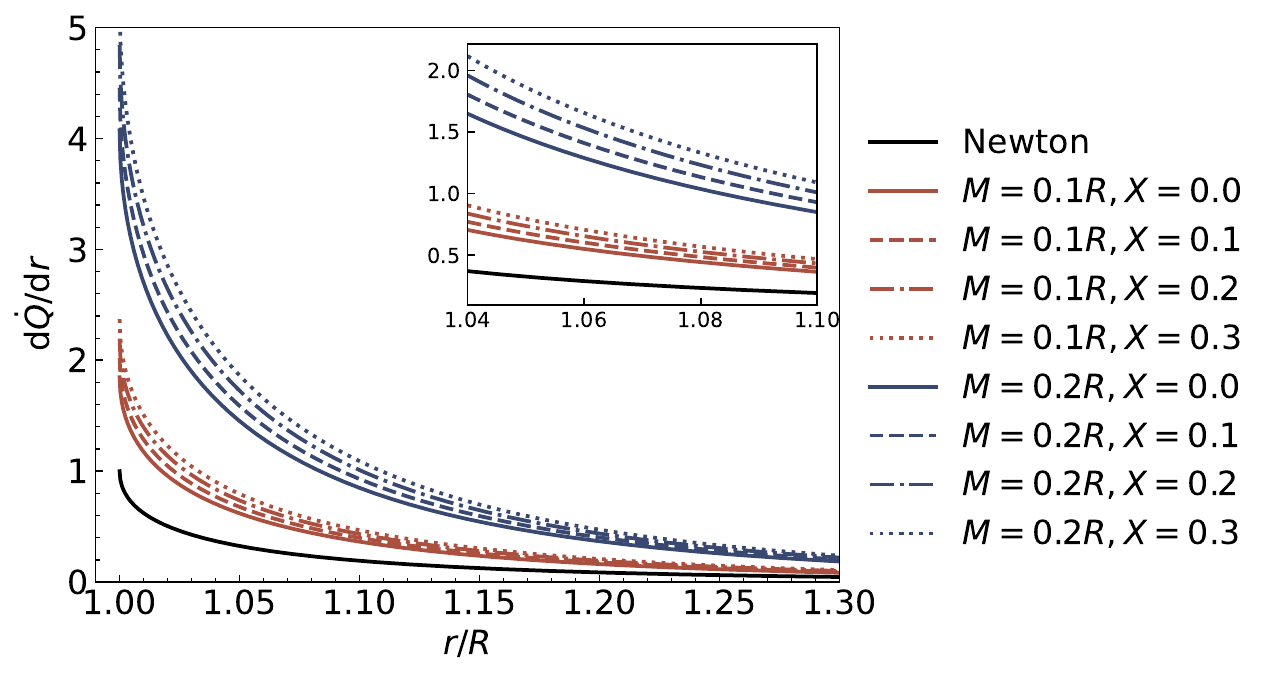}
\caption{The radial dependence of the differential energy deposition rate, $\mathrm{d}\dot{Q}/\mathrm{d}r$, is examined for varying compactness ratios $M/R$. In the Newtonian limit, where the mass parameter vanishes ($M = 0$), the expression simplifies significantly, yielding $\mathrm{d}\dot{Q}/\mathrm{d}r = 1$ exactly at the neutrinosphere radius $r = R$.}
\label{depprate2}
\end{figure*}
 
\begin{table}[h!]
\centering
\caption{The quantity $\dot{Q}$, representing the energy emission rate in $\text{erg/s}$, is determined for different configurations of the compactness ratio $\mathrm{R}/M$ and a range of values for the Lorentz--violating parameter $X$.}
\label{deppratetab}
\begin{tabular}{ccc}
\hline\hline
$X$ & $\mathrm{R}/M$ & $\dot{Q}$ \((\text{erg/s})\) \\
\hline
0.0 & 0 & $1.50 \times 10^{50}$ \\
\hline
\multirow{2}{*}{0.0} & 3 & $4.32 \times 10^{51}$ \\
                      & 4 & $1.10 \times 10^{51}$ \\
\hline
\multirow{2}{*}{0.1} & 3 & $4.73 \times 10^{51}$ \\
                      & 4 & $1.20 \times 10^{51}$ \\
\hline
\multirow{2}{*}{0.2} & 3 & $5.14 \times 10^{51}$ \\
                      & 4 & $1.31 \times 10^{51}$ \\
\hline
\multirow{2}{*}{0.3} & 3 & $5.55 \times 10^{51}$ \\
                      & 4 & $1.41 \times 10^{51}$ \\
\hline\hline
\end{tabular}
\end{table}

The energy deposition rate via neutrino--antineutrino annihilation shows a marked increase within the \textit{metric--affine} framework when compared to the \textit{metric} case \cite{Shi:2025plr}, mainly due to the role played by \textit{non--metricity}, which intensifies the coupling between the neutrino energy flux and the gravitational background, as shown in Figs. \ref{depprate} and \ref{depprate2}. This effect comes from nonlinear curvature deformations that are characteristic of the \textit{metric--affine} geometry, where both the temporal ($\mathrm{g}_{tt}$) and radial ($\mathrm{g}_{rr}$) components of the metric are modified by the Lorentz--violating parameter $X$. On the other hand, in the \textit{metric} formulation, only the radial component $\mathrm{g}_{rr}$ receives a linear correction of the form $(1 + \ell)$. To contextualize our numerical evaluations astrophysically, it is important to note that the compactness ratio $R/M = 4$ effectively models the neutrino-sphere of a massive neutron star (e.g., $M \approx 1.7 M_{\odot}$ and $R \approx 10$ km), which serves as the standard astrophysical site for intense neutrino pair annihilation \cite{Salmonson:1999es,Lambiase:2020iul}. At $\mathrm{R}/M = 4$, the deposition rate in the \textit{metric--affine} setup surpasses the GR benchmark by about $22\%$ (Tab. \ref{deppratetab}) and remains higher than the \textit{metric} result by approximately $4\% \sim 11\%$ in Ref.~\cite{Shi:2025plr}.


\section{Evolution of neutrino phase and its probability conservation}

Throughout the analytical derivation of the neutrino phase in this section, we adopt the standard natural unit system where $c = \hbar = 1$. In this convention, masses and energies share the dimension of energy, while lengths and times carry the dimension of inverse energy, ensuring that the accumulated quantum phase $\chi_k$ is strictly dimensionless.

Taking into account the equations governing the motion of neutrinos in spherically symmetric spacetimes, we begin by adopting a variational formulation based on a Lagrangian formalism. Within the geometric setup defined by the corresponding metric, the procedure involves varying the action with respect to the coordinates describing the particle’s path. Following the methodology proposed in Ref.~\cite{neu18}, this formalism enables a consistent derivation of the motion equations, specifically for the $k$--th eigenmode, while incorporating the curvature effects inherent to the background structure
\begin{align}
\mathcal{L}
& = -\frac{1}{2}  m_{k} \mathrm{g}_{tt}(r)\left(\frac{\mathrm{d}t}{\mathrm{d}\tau}\right)^2-\frac{1}{2}m_{k}\mathrm{g}_{rr}(r)\left(\frac{\mathrm{d}r}{\mathrm{d}\tau}\right)^2 \notag\\
& \quad -\dfrac{1}{2}m_{k}r^2\left(\frac{\mathrm{d}\theta}{\mathrm{d}\tau}\right)^2  
 -\frac{1}{2}m_{k}r^2\sin^2\theta\left(\frac{\mathrm{d}\varphi}{\mathrm{d}\tau}\right)^2.
\end{align}

Setting $\theta = \pi/2$ confines the motion to the equatorial region and considerably reduces the number of non--trivial momentum components. To identify those that remain, one introduces the canonical momentum through the definition $p_\mu = \partial \mathcal{L} / \partial (\mathrm{d}x^\mu/\mathrm{d}\tau)$, where $\tau$ is the affine parameter associated with proper time and $\mathcal{L}$ is the corresponding Lagrangian density. The mass $m_{k}$ refers to the $k$--th neutrino eigenstate under consideration. Once the symmetry condition is enforced, only specific components of $p_\mu$ contribute to the dynamics, and their explicit forms—adapted to this reduced configuration—can be found in Refs.~\cite{Shi:2024flw,neu60}
\begin{align}
p^{(k)t} &= -m_{k}\mathrm{g}_{tt}(r)\frac{\mathrm{d}t}{\mathrm{d}\tau} = -E_{k}, \\
p^{(k)r} &= m_{k}\mathrm{g}_{rr}(r)\frac{\mathrm{d}r}{\mathrm{d}\tau}, \\
p^{(k)\varphi} &= m_{k}r^2\frac{\mathrm{d}\varphi}{\mathrm{d}\tau} = J_{k}.
\end{align}

For a neutrino identified with the $k$--th mass eigenmode, its path through curved spacetime must satisfy the constraint imposed by the mass--shell condition, given by $\mathrm{g}^{\mu\nu} p_\mu p_\nu = -m_k^2$. This equation encodes the fundamental requirement that the particle’s momentum remains compatible with the geometric background, anchoring its motion to the principles of relativistic dynamics. Such consistency between the spacetime curvature and the particle’s four--momentum has been emphasized in prior analyses \cite{neu55,neu54}
\begin{align}
-m_{k}^2 =\mathrm{g}^{tt}p_t^2+\mathrm{g}^{rr}p_r^2+\mathrm{g}^{\varphi\varphi}p_{\varphi}^2.
\end{align}

In the absence of strong gravitational fields, the description of neutrino oscillations is often simplified by adopting the plane wave formalism \cite{neu53,neu54,Shi:2025plr}. Rather than being observed as mass eigenstates, neutrinos appear as linear combinations of such states—referred to as flavor eigenstates—due to the nature of weak interactions governing both their emission and detection. This feature has been thoroughly investigated in previous analyses \cite{neu62,AraujoFilho:2025rzh,neu61,Shi:2025plr,neu63,Shi:2024flw}
\begin{align}
\ket{\nu_{\alpha}} = \sum U_{\alpha i}^{*}\ket{\nu_{i}}.
\end{align}

In the context of neutrino propagation, it is more appropriate to describe their evolution using mass eigenstates $\ket{\nu_i}$, each shaped by a mass--dependent worldline. However, due to the weak interaction mechanism responsible for both production and detection, neutrinos emerge and are measured as flavor states $\nu_\alpha$ with $\alpha = e, \mu, \tau$. These physical states arise from linear combinations of mass eigenstates through a unitary mixing matrix $U$, as formulated in Ref.~\cite{neu41}. To model their motion between emission and detection events, one assigns spacetime points $\left(t_S, \bm{x}_S\right)$ and $\left(t_D, \bm{x}_D\right)$ to the source and detector, respectively. Moreover, each $\ket{\nu_i}$ evolves independently through its own geodesic connecting these respective two points
\begin{align}
\ket{\nu_{i}\left(t_{D},\bm{x}_{D}\right)} = \exp({-\mathbbm{i}\chi_{i}})\ket{\nu_{i}\left(t_{S},\bm{x}_{S}\right)}.
\end{align}
As a consequence of their differing masses, each eigenstate acquires a unique phase during its journey from the emission point to the detector. This accumulated phase is computed using the expression:
\begin{align}
\chi_{i}=\int_{\left(t_{S},\bm{x}_{S}\right)}^{\left(t_{D},\bm{x}_{D}\right)}\mathrm{g}_{\mu\nu}p^{(i)\mu}\mathrm{d}x^{\nu}.
\end{align}

The analysis focuses on flavor transitions that may occur as a neutrino travels from the point of production to the location of detection. Although initially prepared in a specific flavor state $\nu_{\alpha}$, the neutrino can be observed as a different flavor $\nu_{\beta}$ upon arrival. The probability governing this transformation is captured by the following expression:
\begin{align}
\mathcal{P}_{\alpha\beta}
& = |\left\langle \nu_{\beta}|\nu_{\alpha}\left(t_{D}, \bm{x}_{D}\right)\right\rangle|^2 \\
& = \sum_{i,j} U_{\beta i}U_{\beta j}^{*} U_{\alpha j} U_{\alpha i}^{*}\,  \exp{[-\mathbbm{i}(\chi_{i}-\chi_{j})]}.
\end{align}

Assuming this planar configuration ($\theta = \pi/2$), the phase that each neutrino mass eigenstate accumulates through its path is
\begin{align}
\label{fizao}
\chi_{k} & = \int_{\left(t_{S},\bm{x}_{S}\right)}^{\left(t_{D}, \bm{x}_{D}\right)} \mathrm{g}_{\mu\nu} p^{(k)\mu}\mathrm{d}x^{\nu}\notag\\
& = \int_{\left(t_{S},\bm{x}_{S}\right)}^{\left(t_{D}, \bm{x}_{D}\right)}\left[E_{k}\mathrm{d}t - p^{(k)r}\mathrm{d}r-J_{k}\mathrm{d}\varphi\right] \notag\\
& = \pm\frac{m_{k}^2}{2E_0}\int_{r_{S}}^{r_{D}}\sqrt{-\mathrm{g}_{tt}\mathrm{g}_{rr}}\left(1-\dfrac{b^2|\mathrm{g}_{tt}|}{\mathrm{g}_{\varphi\varphi}}\right)^{-\frac{1}{2}}\mathrm{d}r.
\end{align}

In regimes where the gravitational influence remains small—specifically when the ratio $M/r$ is much less than unity—the integrand found in Eq.~(\ref{fizao}) becomes amenable to a series expansion, expressed as follows:
\begin{align}
\label{weak}
&\quad\sqrt{-\mathrm{g}_{tt}\mathrm{g}_{rr}}\left(1-\dfrac{b^2|\mathrm{g}_{tt}|}{\mathrm{g}_{\varphi\varphi}}\right)^{-\frac{1}{2}}\notag\\
&\simeq\dfrac{1}{X_2}\Biggl[\left(1-\frac{b^2}{r^2\sqrt{X_1X_2}}\right)^{-\frac{1}{2}}\notag\\
&\quad-\dfrac{b^2M}{r^3\sqrt{X_1X_2}}\left(1-\frac{b^2}{r^2\sqrt{X_1X_2}}\right)^{-\frac{3}{2}}\Biggr].
\end{align}
It is crucial to clarify the regime of validity for the weak-field expansion applied in Eq.~\eqref{weak}. Although the background geometry corresponds to a black hole solution, our analysis of neutrino phase evolution focuses specifically on the weak gravitational lensing scenario. In this context, the neutrino's distance of closest approach $r_0$ and consequently the impact parameter $b$ is assumed to be significantly larger than the mass of the central compact object ($b \gg M$). Since the radial coordinate of the neutrino satisfies $r \ge r_0$ along its entire trajectory from the source to the detector, the condition $M/r \le M/r_0 \ll 1$ is strictly maintained. This perturbative framework is a well-established standard for evaluating gravitationally induced neutrino oscillation phases in astrophysical lensing environments \cite{Chakrabarty:2023kld,shi2022neutrino,Shi:2024flw,swami2020signature}. As a result, it becomes
\begin{align}
\chi_k&=\dfrac{m_k^2}{2E_0}\dfrac{1}{X_2}\Biggl\{\left(r_D^2-\dfrac{b^2}{\sqrt{X_1X_2}}\right)^{\frac{1}{2}}\notag\\
&\quad-\left(r_S^2-\dfrac{b^2}{\sqrt{X_1X_2}}\right)^{\frac{1}{2}}\notag\\
&\quad+M\Biggl[r_D\left(r_D^2-\dfrac{b^2}{\sqrt{X_1X_2}}\right)^{\frac{1}{2}}\notag\\
&\quad-r_S\left(r_S^2-\dfrac{b^2}{\sqrt{X_1X_2}}\right)^{\frac{1}{2}}\Biggr]\Biggr\}.
\end{align}

The energy of each relativistic neutrino emitted by the source, under this scheme, is characterized by the relation $E_0 = \sqrt{E_k^2 - m_k^2}$, where $E_k$ denotes its total energy. The impact parameter $b$, which quantifies the transverse distance of the trajectory relative to the central object, is defined and discussed in Ref.~\cite{neu18}. As the neutrino traverses the curved spacetime, its path bends inward, reaching a minimum radius $r = r_{0}$. In the weak field regime this closest--approach radius can be computed analytically by applying appropriate perturbative methods to the relevant equations
\begin{align}
\left(\dfrac{\mathrm{d}r}{\mathrm{d}\varphi}\right)_{0} = \pm\dfrac{\mathrm{g}_{\varphi\varphi}}{b^2}\sqrt{\dfrac{1}{-\mathrm{g}_{tt}\mathrm{g}_{rr}}-\dfrac{b^2}{\mathrm{g}_{rr}\mathrm{g}_{\varphi\varphi}}}=0.
\end{align}

The value of $r_0$, representing the nearest radial point reached by the neutrino. This minimum distance is derived by solving the trajectory equation, which becomes tractable through approximations valid when the spacetime curvature remains small
\begin{align}
\label{r0}
r_0 \simeq \dfrac{b}{(X_1X_2)^{\frac{1}{4}}}-M.
\end{align}

Therefore, the phase acquired by a neutrino along its path—from emission to detection, passing through the point of minimal radial distance—is obtained by evaluating the integral for the phase using the approximate form of $r_0$. Incorporating the expression for $r_0$ into the phase calculation leads to the following result
\begin{align}
\label{pphhiii}
&\quad\chi_{k}\left(r_{S}\to r_{0} \to r_{D}\right)\notag\\
&\simeq \frac{{m}_{k}^2}{2E_0}\dfrac{1}{X_2}
\Biggl[\sqrt{r_D^2-r_0^2}+\sqrt{r_S^2-r_0^2}\notag\\
&\quad+M\left(\sqrt{\dfrac{r_D-r_0}{r_D+r_0}}+\sqrt{\dfrac{r_S-r_0}{r_S+r_0}}\right)\Biggr],
\end{align}
which follows
\begin{align}
\chi_{k}
& \simeq \frac{{m}_{k}^2}{2E_0}\dfrac{1}{X_2}
\Biggl[\sqrt{r_D^2-b^2}+\sqrt{r_S^2-b^2}\notag\\
& \quad+M\Biggl(\dfrac{b}{\sqrt{r_D^2-b^2}}+\dfrac{b}{\sqrt{r_S^2-b^2}}\notag\\
&\quad+\sqrt{\dfrac{r_D-b}{r_D+b}}+\sqrt{\dfrac{r_S-b}{r_S+b}}\Biggr)\Biggr].
\end{align}

The next stage consists in performing a series expansion of the derived expression in terms of the small parameter $b/r_{S,D}$, under the assumption that the impact parameter $b$ remains significantly smaller than the radial positions of both the source and the detector. By retaining contributions up to second order, namely $\mathcal{O}(b^2/r_{S,D}^2)$, the approximation simplifies to:
\begin{align}
\chi_k=\dfrac{m_k^2}{2E_0}\dfrac{1}{X_2}(r_D+r_S)\left(1-\dfrac{b^2}{2r_Dr_S}+\dfrac{2M}{r_D+r_S}\right).
\end{align}

The phase acquired by neutrinos diminishes progressively as the Lorentz--violating parameter $X$ grows. For this evaluation, the parameters are chosen as $E_{0} = 10,\mathrm{MeV}$, $r_{D} = 10,\mathrm{km}$, and $r_{S} = 10^5 r_{D}$ to maintain consistency with astrophysical scales.

To accurately describe the flavor transition behavior near the black hole, one must account for the phase variations introduced by different possible paths, each influenced by the curvature--induced deflection \cite{Shi:2025plr,Shi:2024flw,AraujoFilho:2025rzh}
\begin{align}
\Delta\chi_{ij}^{pq}
&= \chi_i^{p}-\chi_j^{q}\notag\\
&= \Delta m_{ij}^2 A_{pq}+\Delta b_{pq}^2 B_{ij},
\end{align}
in which
\begin{align}
\Delta m_{ij}^2 & = m_i^2 - m_j^2,\\
\Delta b_{pq}^2 & = b_{p}^2-b_{q}^2,\\
A_{pq} & = \frac{r_{S} + r_{D}}{2 E_0}\dfrac{1}{X_2}\left(1+\dfrac{2M}{r_D+r_S}-\dfrac{\sum b_{pq}^2}{4r_Dr_S}\right)\notag\\
&\simeq\frac{r_{S} + r_{D}}{2 E_0}\left(1+\dfrac{X}{4}\right)\notag\\
&\quad\times\left(1+\dfrac{2M}{r_D+r_S}-\dfrac{\sum b_{pq}^2}{4r_Dr_S}\right),\\
B_{ij} & = -\frac{\sum m_{ij}^2}{8E_0}\dfrac{1}{X_2}\left(\frac{1}{r_{D}} + \frac{1}{r_{S}}\right)\notag\\
&\simeq-\frac{\sum m_{ij}^2}{8E_0}\left(1+\dfrac{X}{4}\right)\left(\frac{1}{r_{D}} + \frac{1}{r_{S}}\right),\\
\sum b_{pq}^2 & = b_{p}^2 + b_{q}^2,\\
\sum m_{ij}^2 & = m_i^2 + m_j^2.
\end{align}

It is important to note that the notations $\sum b_{pq}^{2}$ and $\sum m_{ij}^{2}$ introduced here are utilized strictly as shorthand macros to denote the simple sum of the two respective squared quantities (i.e., $b_p^2 + b_q^2$ and $m_i^2 + m_j^2$). They are not mathematical summation operators over running indices.

To systematically label the phases corresponding to different neutrino trajectories, the notation $\chi_{i}^{p}$ is adopted, where the superscript $p$ identifies the path determined by a specific impact parameter $b_{p}$. The phase differences that drive oscillation patterns arise from a combination of the neutrino masses $m_{i}$, the mass--squared splittings $\Delta m_{ij}^2$, and the underlying spacetime geometry. When the Lorentz--violating parameter $X$ is set to zero, the resulting expression recovers the standard phase shift formula previously established in Ref.~\cite{neu53}, as one naturally expects.

It is important to note that the term $B_{ij}$ takes into account all contributions related to mass parameters, whereas the Lorentz--violating effects emerge entirely through changes in the coefficient $A_{pq}$, which not only modifies the phase accumulation but also alters the oscillation amplitude.


\section{Neutrino gravitational lensing}

\begin{figure}
    \centering
    \includegraphics[scale=0.45]{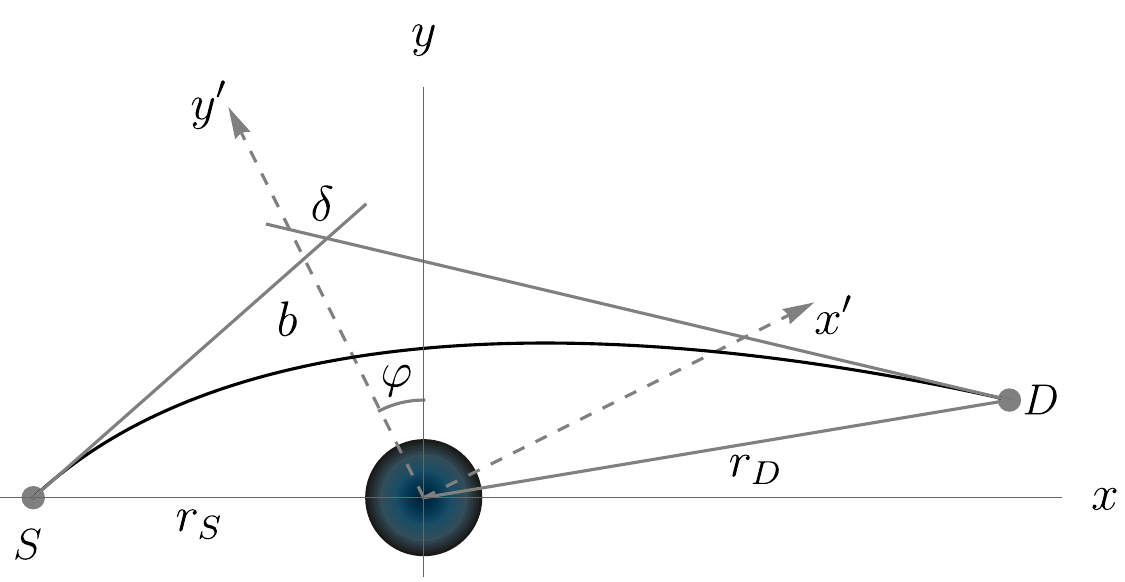}
    \caption{The figure shows the effect of weak lensing on neutrino motion through a curved spacetime geometry. Here, $S$ marks the emission point of the neutrinos, while $D$ identifies the location where they are eventually detected.}
    \label{lenfig}
\end{figure}

In the vicinity of a dense and massive compact object, the curvature of spacetime becomes strong enough to bend neutrino trajectories considerably, steering them away from straight radial paths due to gravitational lensing effects \cite{neu54}. This curvature allows multiple geodesics to emerge from a common source and converge at the same detector location $D$, as shown in Fig.~\ref{lenfig}. Due to these conditions, the usual treatment of a neutrino flavor state must be generalized to account for a coherent sum over all contributing trajectories, each shaped by the spacetime geometry \cite{neu64,neu62,neu65,Shi:2024flw,neu56,neu63,AraujoFilho:2025rzh}:
\begin{align}
|\nu_{\alpha}(t_{D},x_{D})\rangle = N\sum_{i}U_{\alpha i}^{\ast}
\sum_{p} e^{- \mathbbm{i} \chi_{i}^{p}}|\nu_{i}(t_{S}, x_{S})\rangle.
\end{align}

All neutrino trajectories contributing to the detection event are indexed by $p$, with each label identifying a distinct geodesic arriving at the same spatial point. Because the detector receives neutrinos from multiple curved paths, the flavor conversion from an initial state $\nu_{\alpha}$ to a final state $\nu_{\beta}$ emerges through the interference of the corresponding quantum amplitudes. The total probability for this flavor transition is constructed by summing coherently over all such contributions, and takes the form presented below \cite{Shi:2024flw,neu56,neu62,neu63,neu64,neu65}
\begin{align}
\label{lsngkkkk}
\mathcal{P}_{\alpha\beta}^{\mathrm{lensing}} & = |\langle \nu_{\beta}|\nu_{\alpha}(t_{D}, x_{D})\rangle|^{2}\notag\\
& =|N|^{2}\sum_{i, j}U_{\beta i}U_{\beta j}^{\ast}U_{\alpha j}U_{\alpha j}^{\ast}\sum_{p, q}\exp{\left(-\mathbbm{i}\Delta\chi_{ij}^{pq}\right)}.
\end{align}
Following this formulation, the corresponding normalization constant is given by
\begin{align}
|N|^{2} = \left[\sum_{i}|U_{\alpha i}|^{2}\sum_{p,q}\exp\left(-\mathbbm{i}\Delta\chi_{ij}^{pq}\right)\right]^{-1}.
\end{align}

The quantity $\Delta\chi_{ij}^{pq}$, previously introduced, serves as an essential remark in determining the probability of neutrino flavor conversion under the influence of gravitational lensing. This transition probability is shaped by several parameters, including the masses of the neutrino eigenstates, the mass--squared differences $\Delta m_{ij}^2$, and the spacetime geometry generated by the black hole, as shown by Eq.~(\ref{lsngkkkk}). The formalism developed here bears structural resemblance to earlier results obtained in spherically symmetric spacetimes, such as the Schwarzschild geometry \cite{neu53,Shi:2024flw,AraujoFilho:2025rzh,Shi:2025plr}.

We now investigate how gravitational lensing affects neutrino flavor transitions, with particular attention to the role played by the Lorentz--violating parameter $X$. Working within the weak--field regime and assuming a two--flavor system, the flavor transition probability between $\nu_{\alpha}$ and $\nu_{\beta}$ is derived by considering the spatial configuration that includes the emission point, the lensing region, and the detector \cite{neu53,neu54,neu55,Shi:2024flw,neu65}
\begin{align}
\label{gnhjkksdsd}
\mathcal{P}_{\alpha\beta}^{\mathrm{lensing}}
&=\left|N\right|^2\biggl\{2\sum_i\left|U_{\beta i}\right|^2\left|U_{\alpha i}\right|^2\left[1+\cos\left(\Delta b_{12}^2B_{ii}\right)\right]\notag\\
&\quad+\sum_{i\neq j}U_{\beta i}U_{\beta j}^*U_{\alpha j}U_{\alpha i}^*\notag\\
&\quad\times\left[\exp\left(-\mathbbm{i}\Delta m_{ij}^2 A_{11}\right)+\exp\left(-\mathbbm{i}\Delta m_{ij}^2 A_{22}\right)\right]\notag\\
&\quad+\sum_{i\neq j}U_{\beta i}U_{\beta j}^*U_{\alpha j}U_{\alpha i}^*\notag\\
&\quad\times\exp\left(-\mathbbm{i}\Delta b_{12}^2B_{ij}\right)\exp\left(-\mathbbm{i}\Delta m_{ij}^2A_{12}\right)\notag\\
&\quad+\sum_{i\neq j}U_{\beta i}U_{\beta j}^*U_{\alpha j}U_{\alpha i}^*\notag\\
&\quad\times\exp\left(\mathbbm{i}\Delta b_{21}^2B_{ij}\right)\exp\left(-\mathbbm{i}\Delta m_{ij}^2A_{21}\right)\biggr\}.
\end{align}

Eq.~(\ref{gnhjkksdsd}) encodes the probability for neutrino flavor change into a structured sum, where each term is distinguished by specific combinations of mass eigenstates and propagation paths. When $i = j$, the expression accounts for the isolated evolution of a single mass eigenstate without any interference. In contrast, the case $i \neq j$ with $p = q$ represents quantum interference between different mass states traveling along the same geodesic, where distinct phases are accumulated due to their mass differences.

Interference becomes more pronounced in situations where both mass indices and trajectory labels differ, that is, $i \neq j$ and $p \neq q$. In addition, these cross contributions are treated separately for $p < q$ and $p > q$ to reflect the asymmetry in the accumulated phases, which arises from variations in the path lengths and the curvature effects associated with each ``route''.

In the two-flavor approximation, the theoretical framework becomes substantially simpler. Within this setting, the connection between flavor states and mass eigenstates is established through a $2 \times 2$ unitary matrix, which is entirely determined by a single parameter—the mixing angle $\alpha$ \cite{neu43}
\begin{align}
\label{matrixU}
U\equiv\left(\begin{matrix}
\cos\alpha&\sin\alpha\\
-\sin\alpha&\cos\alpha
\end{matrix}\right).
\end{align}

Replacing the mixing matrix defined earlier with its explicit form yields a specific expression for the probability of flavor conversion from $\nu_e$ to $\nu_\mu$ when applied to the general oscillation framework
\begin{align}
\label{gbabil}
\mathcal{P}_{\alpha\beta}^{\mathrm{lensing}}
&=\left|N\right|^2\sin^{2}2\alpha\notag\\
&\quad\times\biggl[\sin^2\left(\dfrac{1}{2}\Delta m_{12}^2A_{11}\right)+\sin^2\left(\dfrac{1}{2}\Delta m_{12}^2A_{22}\right)\notag\\
&\quad+\dfrac{1}{2}\cos\left(\Delta b_{12}^2B_{11}\right)+\dfrac{1}{2}\cos\left(\Delta b_{12}^2B_{22}\right)\notag\\
&\quad-\cos\left(\Delta b_{12}^2B_{12}\right)\cos\left(\Delta m_{12}^2A_{12}\right)\biggr].
\end{align}

Taking into account the structure of the leptonic mixing matrix given in Eq.~(\ref{matrixU}) and incorporating the phase shifts associated with each neutrino path, the normalization factor can be expressed as follows:
\begin{align}
\left|N\right|^2&=\biggl[2+2\cos^2\alpha\cos\left(\Delta b_{12}^2B_{11}\right)\notag\\
&\quad+2\sin^2\alpha\cos\left(\Delta b_{12}^2B_{22}\right)\biggr]^{-1}.
\end{align}


\section{Numerical estimations}

To study neutrino oscillations in the curved geometry produced by the black hole in question, it becomes necessary to evaluate the lensing probabilities provided in Eq.~(\ref{gbabil}). In the original coordinate system $(x, y)$, the lens occupies the central position, while the source and detector are situated at distances $r_S$ and $r_D$ from the origin. For computational clarity, a rotated frame $(x', y')$ is introduced by applying a transformation through an angle $\varphi$. This change of coordinates gives rise to the following relation \cite{neu53,Shi:2024flw,Shi:2025plr}
\begin{align}
x' = x\cos\varphi + y\sin\varphi, \quad y' = -x\sin\varphi + y\cos\varphi .
\nonumber
\end{align}

Setting $\varphi = 0$ leads to a particularly symmetric setup in which the source, lens, and detector are perfectly aligned along a straight line within the plane. This colinear arrangement places all three elements on the same axis, simplifying the analysis of how neutrinos propagate through the gravitational field.

As detailed in Refs.~\cite{neu53,Shi:2024flw,Shi:2025plr}, the deflection experienced by neutrinos due to spacetime curvature—characterized by the angle $\delta$—is governed by the impact parameter $b$. The relationship between these quantities is described by the following expression \cite{14}:
\begin{align}
\label{delta}
\delta \sim\frac{y_{D}'- b}{x_{D}'}=-\dfrac{3\pi X}{8}-\dfrac{4M_X}{b}.
\end{align}
with
\begin{align}
M_X = M+\dfrac{3X}{8}
\end{align}

Placing the detector at coordinates $(x_{D}', y_{D}')$ within the rotated reference frame and making use of the relation $\sin\varphi = b/r_S$, one can rewritten the deflection angle originally given in Eq.~(\ref{delta}) into the following alternative form:
\begin{align}
\label{solve_b}
&\quad\left(4M_Xx_D+y_Db+\dfrac{3\pi X x_D b}{9}\right)\sqrt{1-\dfrac{b^2}{r_S^2}}\notag\\
&=b^2\left(\dfrac{x_D}{r_S}+1-\dfrac{3\pi X y_D}{8r_S}\right)-\dfrac{4M_X y_D b}{r_S}.
\end{align}

To systematically evaluate the lensing probabilities and the phase shifts, we model a physically viable astrophysical scenario wherein a high-energy neutrino source is located behind a massive lens. The detector is placed at a distance $r_D = 10^8 \, \mathrm{km}$, and the source is located at $r_S = 10^5 r_D$. For each azimuthal angle $\varphi$, the lensing equation is numerically solved to obtain the two real roots representing the impact parameters, $b_1$ and $b_2$. The full set of initial conditions, including the typical neutrino mass-squared difference $|\Delta m^2| = 10^{-3} \, \mathrm{eV^2}$ and energy $E_0 = 10 \, \mathrm{MeV}$, is summarized in Table \ref{tab:numerical_parameters}.

\begin{table}[htpb]
\centering
\caption{Summary of the phenomenological and orbital parameters used in the numerical evaluation of neutrino lensing and oscillations.}
\label{tab:numerical_parameters}
\renewcommand{\arraystretch}{1.3}
\resizebox{\columnwidth}{!}{
\begin{tabular}{lcc}
\hline\hline
\textbf{Physical Quantity} & \textbf{Symbol} & \textbf{Value / Definition} \\
\hline
Neutrino energy & $E_0$ & $10 \, \mathrm{MeV}$ \\
Detector distance & $r_D$ & $10^8 \, \mathrm{km}$ \\
Source distance & $r_S$ & $10^{13} \, \mathrm{km}$ \\
Mass-squared difference & $|\Delta m^2|$ & $10^{-3} \, \mathrm{eV^2}$ \\
Lightest neutrino mass & $m_1$ & $0.00, 0.01, 0.02 \, \mathrm{eV}$ \\
Mixing angle & $\alpha$ & $\pi/5, \, \pi/6$ \\
\hline\hline
\end{tabular}
}
\end{table}

\begin{figure*}
\centering
\includegraphics[height=5cm]{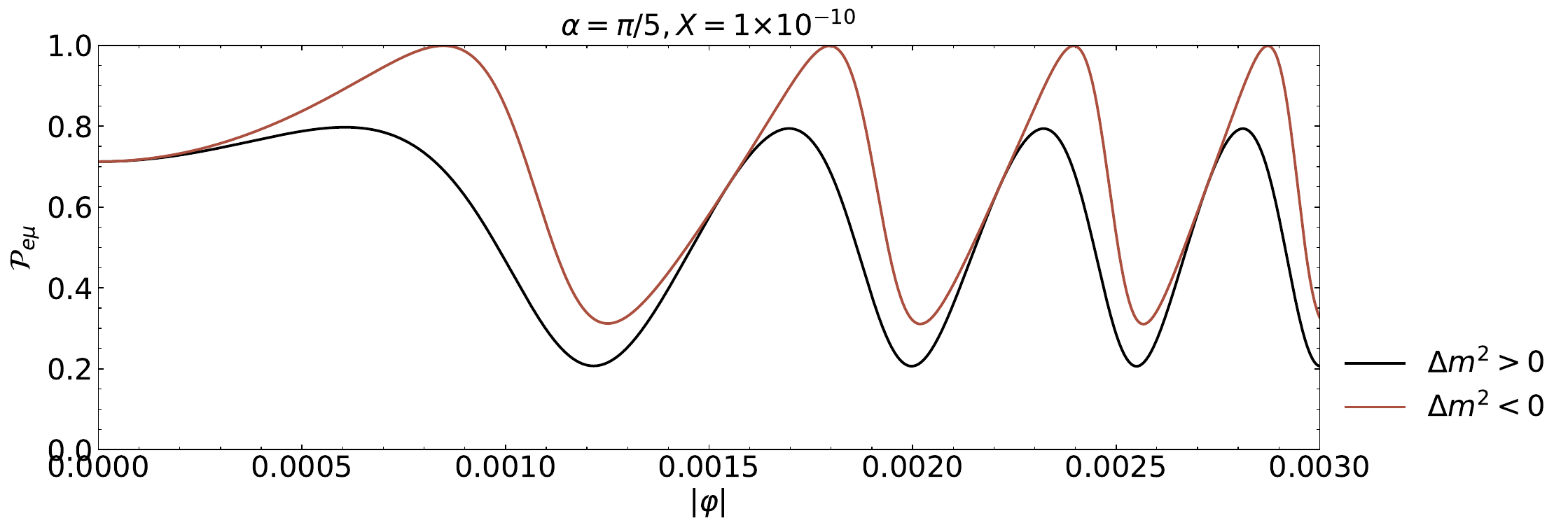}
\includegraphics[height=5cm]{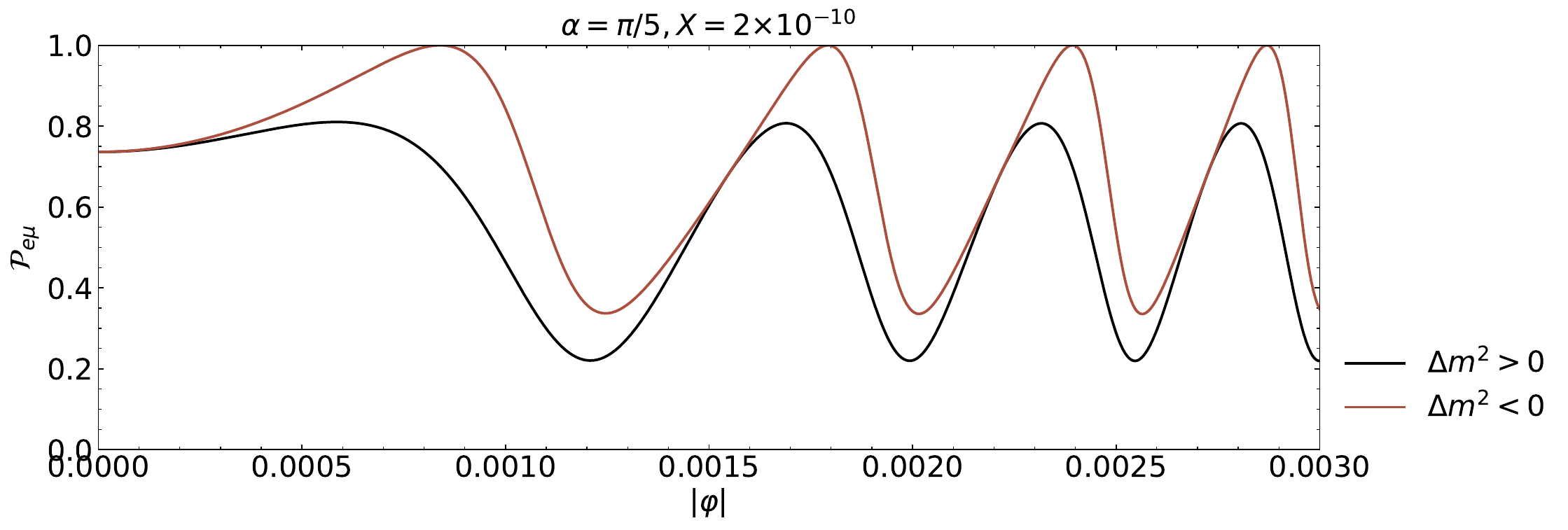}
\includegraphics[height=5cm]{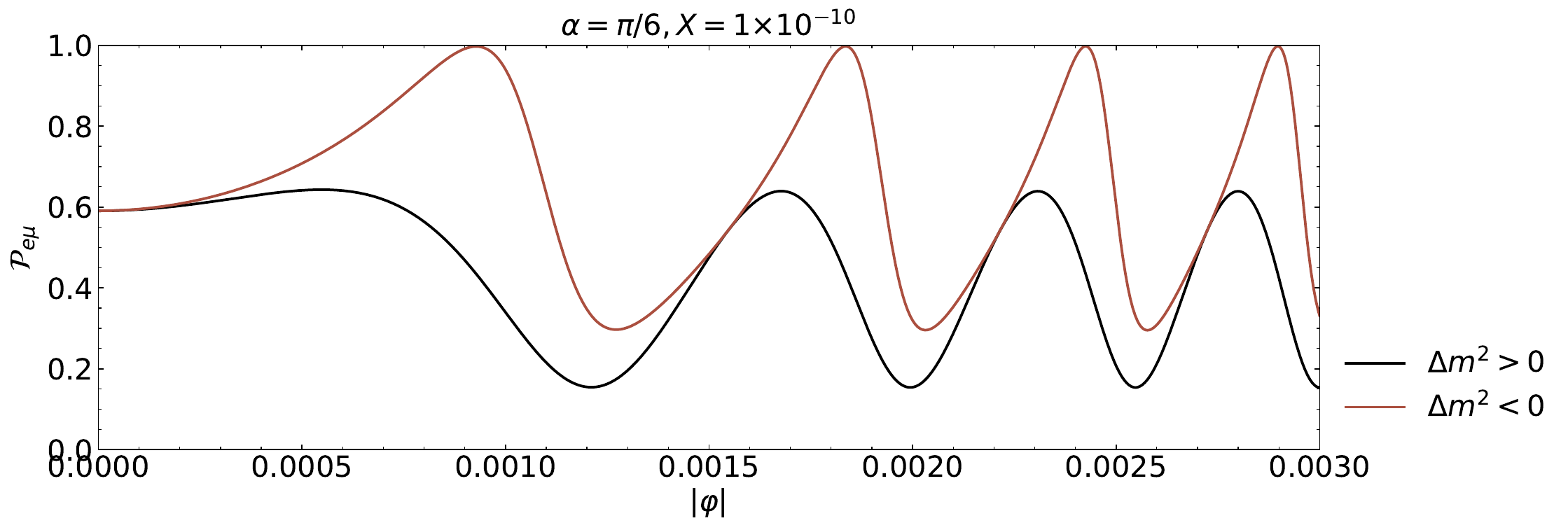}
\includegraphics[height=5cm]{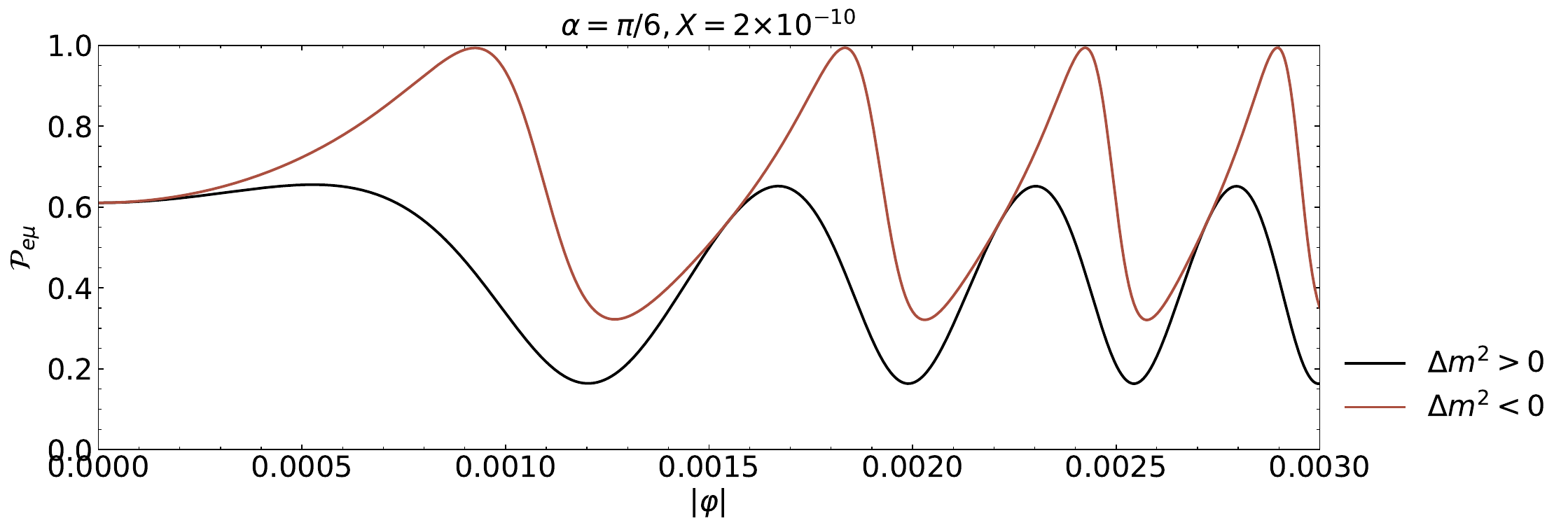}
\caption{\label{fig:prob1} The transition probability for $\nu_e \to \nu_\mu$ is analyzed as a function of the azimuthal angle $\varphi$ for two values of the Lorentz--violating parameter, $X = 1\times10^{-10}$ and $X = 3\times10^{-10}$. The study is performed in the two--flavor approximation, considering both normal and inverted mass orderings. Additionally, mixing angles $\alpha = \pi/5$ and $\alpha = \pi/6$ are used to assess the sensitivity of the oscillation pattern to variations in the mixing parameter.}
\end{figure*}

The behavior of $\nu_e \to \nu_\mu$ oscillations in a Lorentz--violating background, analyzed within the \textit{metric--affine} formalism, displays a marked sensitivity to neutrino mass ordering. As shown in Fig.~\ref{fig:prob1}, the normal hierarchy ($\Delta m^2 > 0$) consistently yields lower transition probabilities than the inverted one ($\Delta m^2 < 0$) for small curvature angles $\varphi \in [0, 0.003]$, indicating a stronger curvature coupling for the inverted spectrum under \textit{non--metricity} effects.

\begin{figure*}
\centering
\includegraphics[height=5cm]{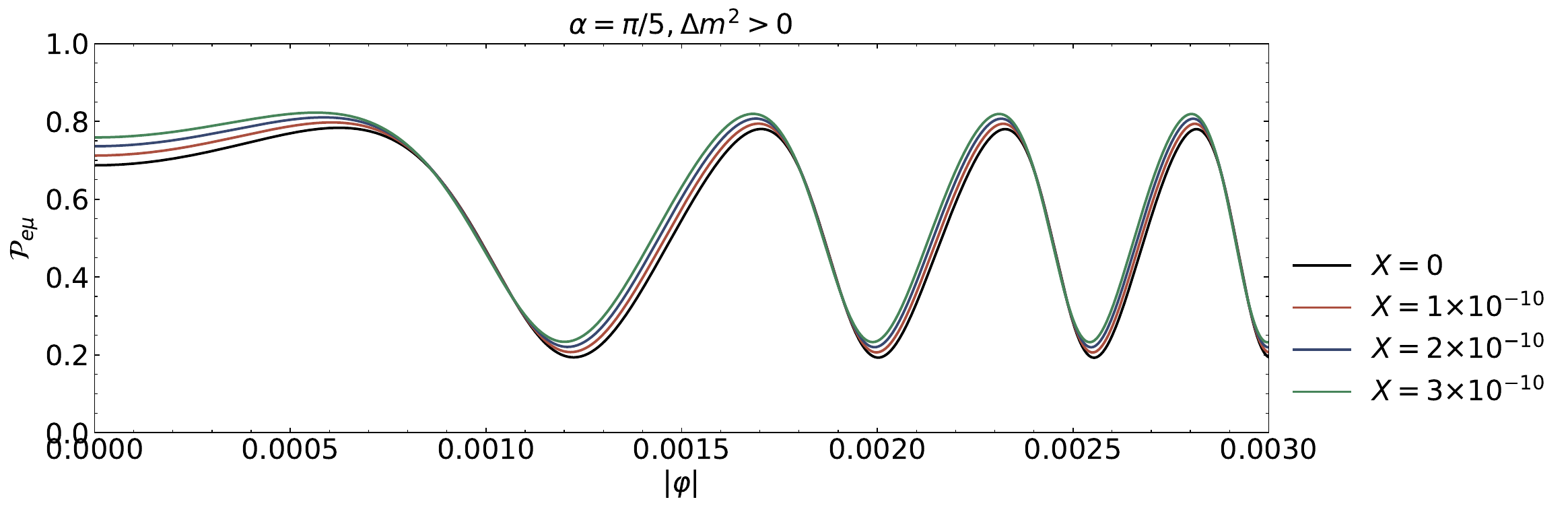}
\includegraphics[height=5cm]{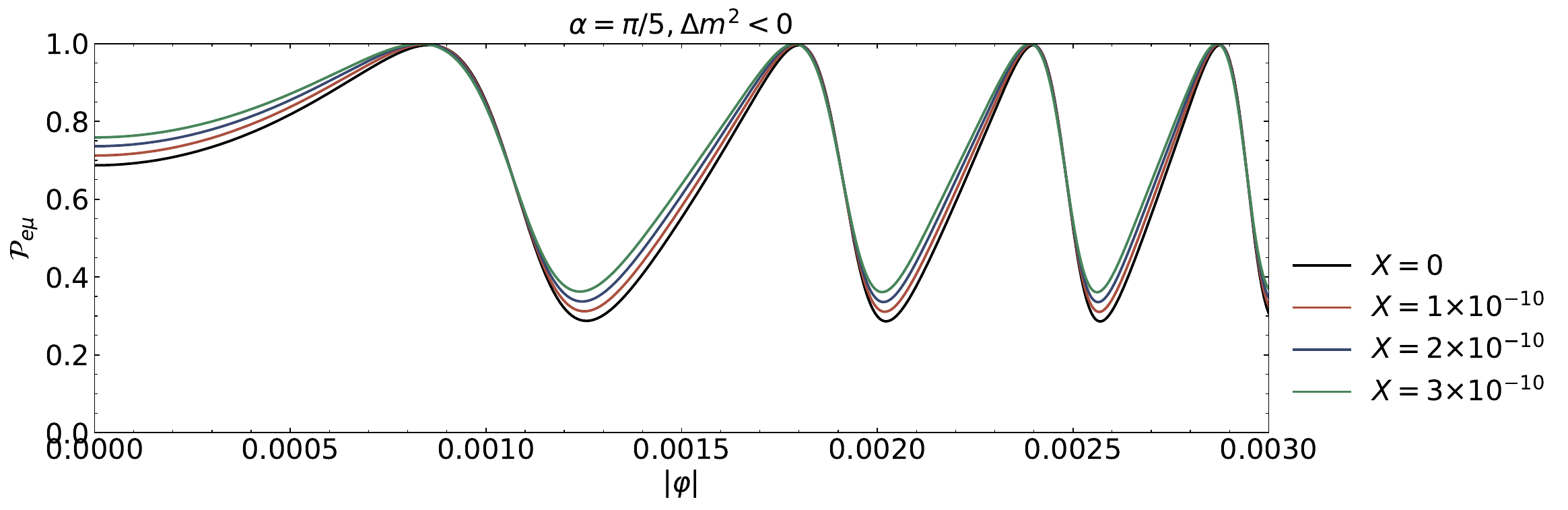}
\includegraphics[height=5cm]{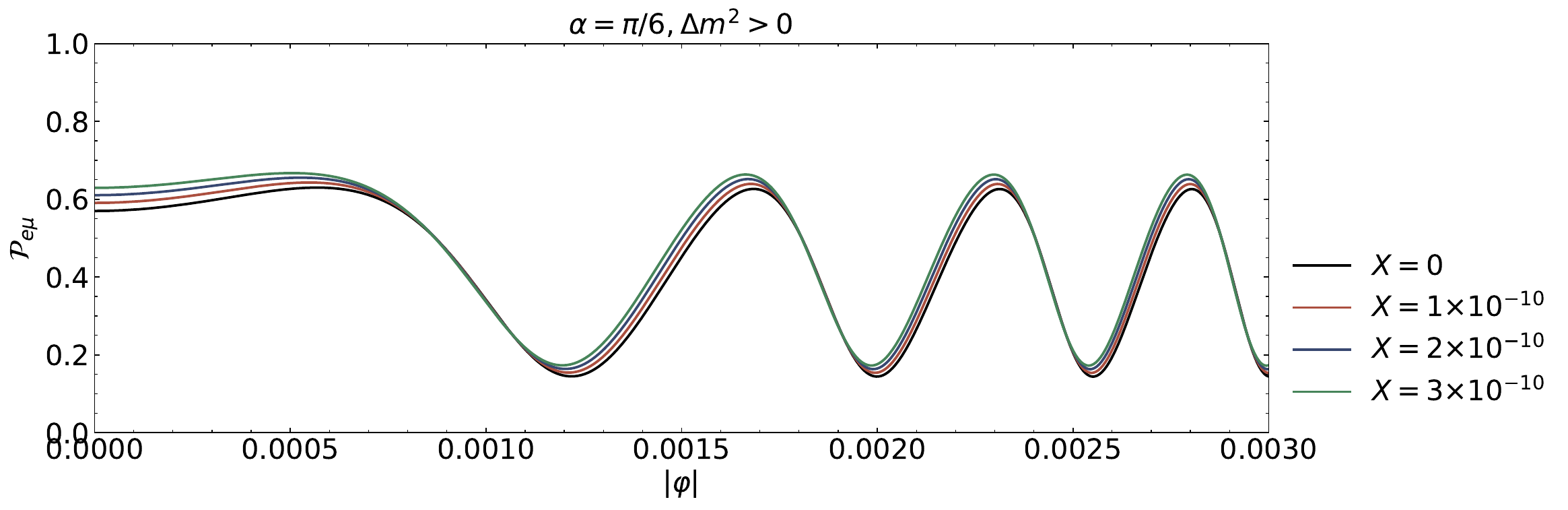}
\includegraphics[height=5cm]{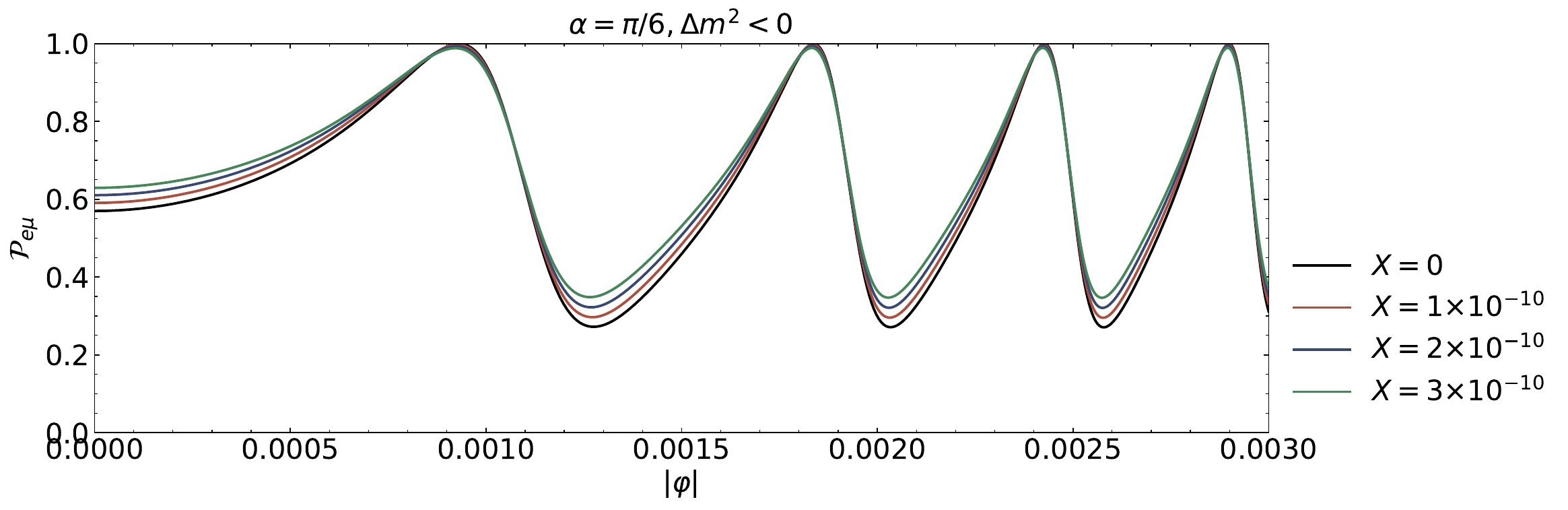}
\caption{\label{fig:prob2} The $\nu_e \to \nu_\mu$ conversion probability is evaluated as a function of the azimuthal angle $\varphi$ for $X = 0$, $1\times10^{-10}$, $2\times10^{-10}$, and $3\times10^{-10}$, considering both normal and inverted orderings in a two--flavor scheme. The impact of the mixing angle is analyzed for $\alpha = \pi/5$ and $\alpha = \pi/6$.}
\end{figure*}

\begin{figure*}
\centering
\includegraphics[height=5cm]{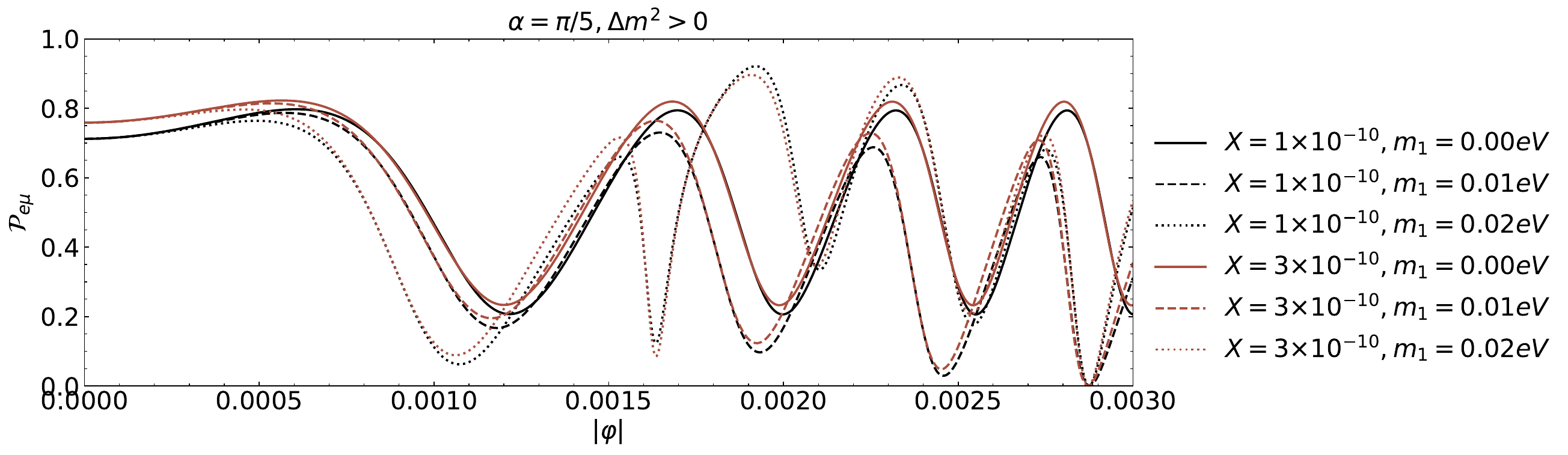}
\includegraphics[height=5cm]{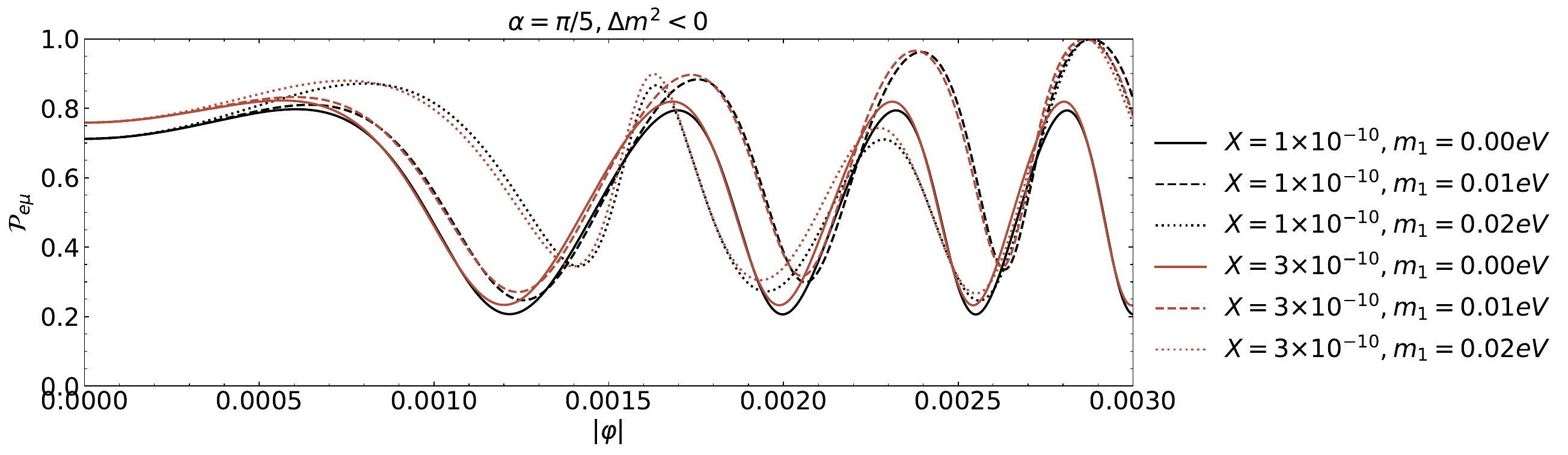}
\includegraphics[height=5cm]{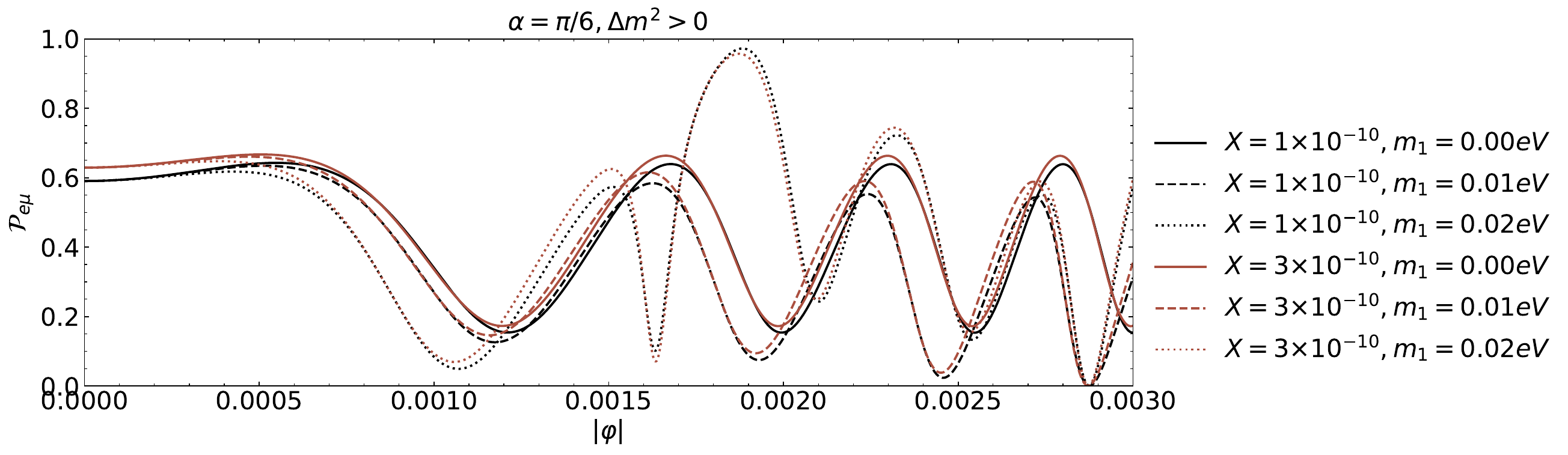}
\includegraphics[height=5cm]{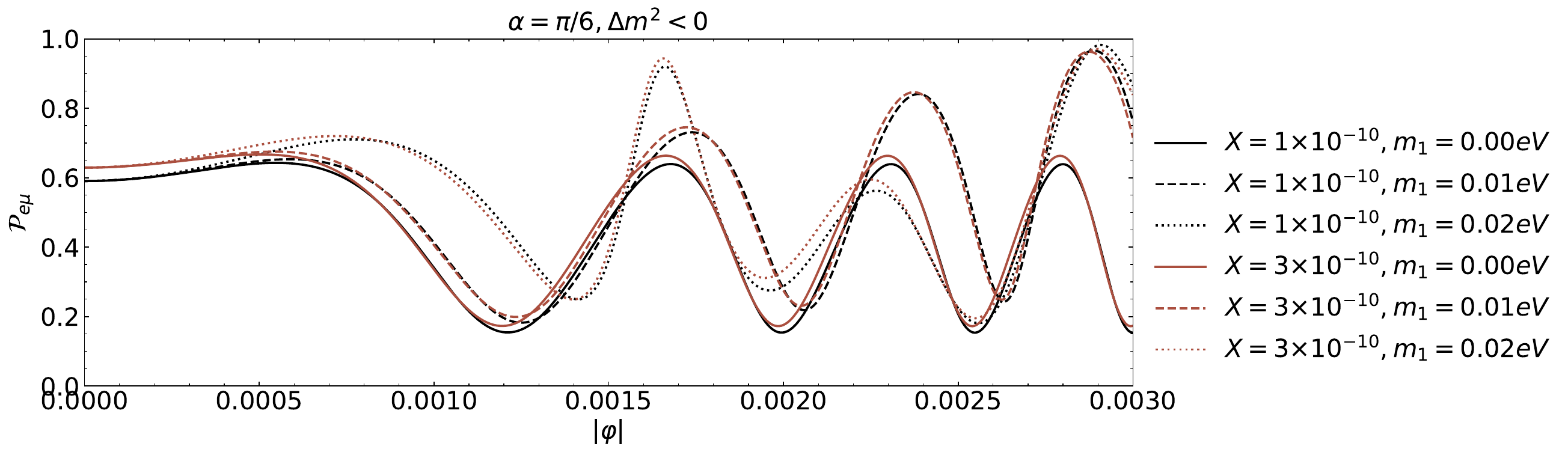}
\caption{\label{fig:prob3} Neutrino oscillation probabilities are shown as functions of the azimuthal angle $\varphi$ for both normal and inverted hierarchies. Lorentz violation is illustrated using $X = 1\times10^{-10}$ (black) and $X = 3\times10^{-10}$ (red). Line styles indicate the lightest neutrino mass: solid for $m_1 = 0,\mathrm{eV}$, dashed for $0.01,\mathrm{eV}$, and dotted for $0.02,\mathrm{eV}$.}
\end{figure*}

The Lorentz--violating parameter $X$ influences oscillation dynamics in two main ways: Fig.\ref{fig:prob2} shows that increasing $X$ enhances the transition probability, while Fig.\ref{fig:prob3} reveals a nonlinear growth in oscillation amplitude, signaling an intensified interference among mass eigenstates. As the lightest mass $m_1$ increases, the amplitude in the normal ordering is suppressed, suggesting that interactions between $m_1$ and curvature inhibit flavor conversion. Interestingly, while $X$ and $m_1$ modulate the amplitude, the oscillation frequency remains unchanged—controlled predominantly by global spacetime curvature.

These effects result from multi--scale interactions inherent to the \textit{metric--affine} geometry. Time--like geodesic convergence is altered by curvature terms arising from independent connection variables, modifying phase accumulation. Moreover, the Bumblebee field enhances the coupling between gravity and the mass basis, amplifying $X$--driven interference. The inclusion of $m_1$ introduces an inertial contribution that partially offsets these effects, reflecting deviations from the weak equivalence principle.

In contrast to the \textit{metric} case, where Lorentz--violating corrections through the parameter $\ell$ only suppress transition probabilities and shifted peaks via the mixing angle $\alpha$, the \textit{metric--affine} model allows $X$ to boost both the probability and amplitude, due to direct coupling between the affine structure and \textit{non--metric} curvature. Furthermore, while both frameworks register suppression of normal ordering with rising $m_1$, only the affine setup reflects a significant enhancement of the inverted ordering, attributed to competitive mass--curvature interactions.

\section{Conclusion}

This study assessed the consequences of spontaneous Lorentz symmetry breaking in bumblebee gravity on neutrino behavior, focusing specifically on the modifications induced \textit{by non--metricity}. By analyzing the energy deposition from the process $\nu \bar{\nu} \to e^+ e^-$, it was shown that the parameter $X$ amplified the total emission rate beyond the predictions of both general relativity and the purely \textit{metric} bumblebee model. For instance, at a representative compactness of $R/M = 4$, the energy output surpassed the GR baseline by approximately $22\%$, and exceeded the \textit{metric} case by up to $11\%$.

In contrast to the \textit{metric} formulation, where only $\mathrm{g}_{rr}$ was linearly corrected by the Lorentz--violating term $\ell$, the \textit{metric--affine} geometry allowed a nonlinear coupling that significantly strengthens the gravitational interaction with neutrino energy fluxes. These modifications also altered the oscillation phase: larger values of $X$ increased the path length, thus enhancing the transition probability and oscillation amplitude. Numerical results showed that while the inverted mass orderings consistently yielded stronger conversion than the normal one, only in the \textit{affine} framework did both the amplitude and the probability grew with increasing $X$.

Therefore, in comparison to the metric formulation, the presence of \textit{non--metricity} in bumblebee gravity resulted in stronger flavor conversion effects, enhanced energy deposition, and more pronounced sensitivity to mass ordering---especially in the inverted scenario. As a further perspective, it would be worthwhile to explore the influence of black holes arising in both \textit{metric} and \textit{metric--affine} gravity on Unruh--DeWitt effects, through the lens of the recent study presented in Ref.~\cite{Barros:2025din,Barros:2024wdv}.


\section*{Acknowledgments}
\hspace{0.5cm} A. A. Araújo Filho is supported by Conselho Nacional de Desenvolvimento Cient\'{\i}fico e Tecnol\'{o}gico (CNPq) and Fundação de Apoio à Pesquisa do Estado da Paraíba (FAPESQ), project No. 150891/2023-7.

\bibliographystyle{ieeetr}
\bibliography{main}

\begin{thebibliography}{10}

\bibitem{filho2023vacuum}
A.~A. Ara{\'u}jo~Filho, J.~R. Nascimento, A.~Y. Petrov, and P.~J.
  Porf{\'\i}rio, ``Vacuum solution within a metric-affine bumblebee gravity,''
  {\em Physical Review D}, vol.~108, no.~8, p.~085010, 2023.

\bibitem{bluhm2006overview}
R.~Bluhm, ``Overview of the standard model extension: implications and
  phenomenology of lorentz violation,'' in {\em Special Relativity: Will it
  Survive the Next 101 Years?}, pp.~191--226, Springer, 2006.

\bibitem{colladay1998lorentz}
D.~Colladay and V.~A. Kosteleck{\`y}, ``Lorentz-violating extension of the
  standard model,'' {\em Physical Review D}, vol.~58, no.~11, p.~116002, 1998.

\bibitem{bluhm2008spontaneous}
R.~Bluhm, S.-H. Fung, and V.~A. Kosteleck{\`y}, ``Spontaneous lorentz and
  diffeomorphism violation, massive modes, and gravity,'' {\em Physical Review
  D}, vol.~77, no.~6, p.~065020, 2008.

\bibitem{kostelecky2004gravity}
V.~A. Kosteleck{\`y}, ``Gravity, lorentz violation, and the standard model,''
  {\em Physical Review D}, vol.~69, no.~10, p.~105009, 2004.

\bibitem{3}
P.~Horava, ``Quantum gravity at a lifshitz point,'' {\em Phys. Rev.}, vol.~79,
  p.~084008, 2009.

\bibitem{5}
T.~Jacobson and D.~Mattingly, ``Gravity with a dynamical preferred frame,''
  {\em Phys. Rev. D}, vol.~64, p.~024028, 2001.

\bibitem{4}
S.~Carroll, J.~Harvey, V.~Kostelecky, C.~Lane, and T.~Okamoto, ``Noncommutative
  field theory and lorentz violation,'' {\em Phys. Rev. Lett.}, vol.~87,
  p.~141601, 2001.

\bibitem{7}
G.~Bengochea and R.~Ferraro, ``Dark torsion as the cosmic speed-up,'' {\em
  Phys. Rev. D}, vol.~79, p.~124019, 2009.

\bibitem{1}
V.~Kostelecky and S.~Samuel, ``Spontaneous breaking of lorentz symmetry in
  string theory,'' {\em Phys. Rev. D}, vol.~39, p.~683, 1989.

\bibitem{8}
A.~Cohen and S.~Glashow, ``Very special relativity,'' {\em Phys. Rev. Lett.},
  vol.~97, p.~021601, 2006.

\bibitem{ghosh2023does}
R.~Ghosh, S.~Nair, L.~Pathak, S.~Sarkar, and A.~S. Sengupta, ``Does the speed
  of gravitational waves depend on the source velocity?,'' {\em Physical Review
  D}, vol.~108, no.~12, p.~124017, 2023.

\bibitem{2}
J.~Alfaro, H.~Morales-Tecotl, and L.~Urrutia, ``Loop quantum gravity and light
  propagation,'' {\em Phys. Rev. D}, vol.~65, p.~103509, 2002.

\bibitem{6}
S.~Dubovsky, P.~Tinyakov, and I.~Tkachev, ``Massive graviton as a testable cold
  dark matter candidate,'' {\em Phys. Rev. Lett.}, vol.~94, p.~181102, 2005.

\bibitem{11}
V.~Kostelecky and S.~Samuel, ``Phenomenological gravitational constraints on
  strings and higher dimensional theories,'' {\em Phys. Rev. Lett.}, vol.~63,
  p.~224, 1989.

\bibitem{AraujoFilho:2024ykw}
A.~A. Ara{\'u}jo~Filho, J.~R. Nascimento, A.~Y. Petrov, and P.~Porf{\'\i}rio,
  ``An exact stationary axisymmetric vacuum solution within a metric-affine
  bumblebee gravity,'' {\em Journal of Cosmology and Astroparticle Physics},
  vol.~2024, no.~07, p.~004, 2024.

\bibitem{13}
R.~Bluhm, N.~Gagne, R.~Potting, and A.~Vrublevskis, ``Constraints and stability
  in vector theories with spontaneous lorentz violation,'' {\em Phys. Rev. D},
  vol.~77, p.~125007, 2008.

\bibitem{KhodadiPoDU2023}
M.~Khodadi and M.~Schreck, ``Hubble tension as a guide for refining the early
  universe: Cosmologies with explicit local lorentz and diffeomorphism
  violation,'' {\em Physics of the Dark Universe}, vol.~39, p.~101170, 2023.

\bibitem{liu2024shadow}
W.~Liu, D.~Wu, and J.~Wang, ``Shadow of slowly rotating kalb-ramond black
  holes,'' {\em arXiv preprint arXiv:2407.07416}, 2024.

\bibitem{llvv2}
A.~Ara{\'u}jo~Filho, H.~Hassanabadi, J.~Reis, and L.~Lisboa-Santos,
  ``Thermodynamics of a quantum ring modified by lorentz violation,'' {\em
  Physica Scripta}, vol.~98, no.~6, p.~065943, 2023.

\bibitem{Magalhaes:2025nql}
R.~B. Magalh{\~a}es, L.~A. Lessa, and M.~M. Ferreira~Jr, ``Wormholes in
  lorentz-violating gravity,'' {\em arXiv preprint arXiv:2505.07590}, 2025.

\bibitem{12}
Q.~Bailey and V.~Kostelecky, ``Signals for lorentz violation in post-newtonian
  gravity,'' {\em Phys. Rev. D}, vol.~74, p.~045001, 2006.

\bibitem{10}
V.~Kostelecky and S.~Samuel, ``Gravitational phenomenology in higher
  dimensional theories and strings,'' {\em Phys. Rev. D}, vol.~40, p.~1886,
  1989.

\bibitem{9}
V.~Kostelecky, ``Gravity, lorentz violation, and the standard model,'' {\em
  Phys. Rev. D}, vol.~69, p.~105009, 2004.

\bibitem{Liu:2024wpa}
W.~Liu, C.~Wen, and J.~Wang, ``{Lorentz violation alleviates gravitationally
  induced entanglement degradation},'' {\em JHEP}, vol.~01, p.~184, 2025.

\bibitem{llvv1}
K.~M. Amarilo, M.~Ferreira~Filho, A.~A. Ara{\'u}jo~Filho, and J.~A. A.~S. Reis,
  ``Gravitational waves effects in a lorentz--violating scenario,'' {\em
  Physics Letters B}, vol.~855, p.~138785, 2024.

\bibitem{Liu:2024oas}
W.~Liu, D.~Wu, and J.~Wang, ``{Static neutral black holes in Kalb-Ramond
  gravity},'' {\em JCAP}, vol.~09, p.~017, 2024.

\bibitem{yang2023static}
K.~Yang, Y.-Z. Chen, Z.-Q. Duan, and J.-Y. Zhao, ``Static and spherically
  symmetric black holes in gravity with a background kalb-ramond field,'' {\em
  Physical Review D}, vol.~108, no.~12, p.~124004, 2023.

\bibitem{heidari2024scattering}
N.~Heidari, C.~F. Macedo, A.~A. Ara{\'u}jo~Filho, and H.~Hassanabadi,
  ``Scattering effects of bumblebee gravity in metric-affine formalism,'' {\em
  The European Physical Journal C}, vol.~84, no.~11, p.~1221, 2024.

\bibitem{anacleto2018lorentz}
M.~Anacleto, F.~Brito, E.~Maciel, A.~Mohammadi, E.~Passos, W.~Santos, and
  J.~Santos, ``Lorentz-violating dimension-five operator contribution to the
  black body radiation,'' {\em Physics Letters B}, vol.~785, pp.~191--196,
  2018.

\bibitem{araujo2022thermal}
A.~A. Ara{\'u}jo~Filho, {\em Thermal aspects of field theories}.
\newblock Amazon. com, 2022.

\bibitem{aa2021lorentz}
A.~AA~Filho, ``Lorentz-violating scenarios in a thermal reservoir,'' {\em The
  European Physical Journal Plus}, vol.~136, no.~4, pp.~1--14, 2021.

\bibitem{araujo2021thermodynamic}
A.~A. Ara{\'u}jo~Filho and R.~V. Maluf, ``Thermodynamic properties in
  higher-derivative electrodynamics,'' {\em Brazilian Journal of Physics},
  vol.~51, no.~3, pp.~820--830, 2021.

\bibitem{araujo2021higher}
A.~Ara{\'u}jo~Filho and A.~Y. Petrov, ``Higher-derivative lorentz-breaking
  dispersion relations: a thermal description,'' {\em The European Physical
  Journal C}, vol.~81, no.~9, p.~843, 2021.

\bibitem{reis2021thermal}
J.~Reis {\em et~al.}, ``Thermal aspects of interacting quantum gases in
  lorentz-violating scenarios,'' {\em The European Physical Journal Plus},
  vol.~136, no.~3, pp.~1--30, 2021.

\bibitem{araujo2022does}
A.~A. Ara{\'u}jo~Filho and J.~Reis, ``How does geometry affect quantum
  gases?,'' {\em International Journal of Modern Physics A}, vol.~37,
  no.~11n12, p.~2250071, 2022.

\bibitem{paperrainbow}
A.~A. Araújo~Filho, J.~Furtado, H.~Hassanabadi, and J.~Reis, ``Thermal
  analysis of photon--like particles in rainbow gravity,'' {\em Physics of Dark
  Universe}, vol.~42, no.~8, p.~101310, 2023.

\bibitem{14}
R.~Casana, A.~Cavalcante, F.~Poulis, and E.~Santos, ``Exact schwarzschild-like
  solution in a bumblebee gravity model,'' {\em Phys. Rev. D}, vol.~97,
  p.~104001, 2018.

\bibitem{kanzi2019gup}
S.~Kanzi and I.~Sakall{\i}, ``Gup modified hawking radiation in bumblebee
  gravity,'' {\em Nuclear Physics B}, vol.~946, p.~114703, 2019.

\bibitem{Khodadi:2025wuw}
M.~Khodadi, G.~Lambiase, L.~Mastrototaro, and T.~K. Poddar, ``Primordial
  gravitational waves from spontaneous lorentz symmetry breaking,'' {\em
  Physics Letters B}, vol.~867, p.~139597, 2025.

\bibitem{Khodadi:2022mzt}
M.~Khodadi, G.~Lambiase, and A.~Sheykhi, ``{Constraining the Lorentz-violating
  bumblebee vector field with big bang nucleosynthesis and gravitational
  baryogenesis},'' {\em Eur. Phys. J. C}, vol.~83, no.~5, p.~386, 2023.

\bibitem{15}
A.~Ovgun, K.~Jusufi, and I.~Sakalli, ``Gravitational lensing under the effect
  of weyl and bumblebee gravities: Applications of gauss-bonnet theorem,'' {\em
  Annals Phys.}, vol.~399, p.~193, 2018.

\bibitem{18}
Z.~Cai and R.-J. Yang, ``Accretion of the vlasov gas onto a schwarzschild-like
  black hole,'' {\em Physics of the Dark Universe}, vol.~42, p.~101292, 2023.

\bibitem{17}
R.-J. Yang, H.~Gao, Y.~Zheng, and Q.~Wu, ``Effects of lorentz breaking on the
  accretion onto a schwarzschild-like black hole,'' {\em Commun. Theor. Phys.},
  vol.~71, p.~568, 2019.

\bibitem{19}
R.~Oliveira, D.~Dantas, and C.~Almeida, ``Quasinormal frequencies for a black
  hole in a bumblebee gravity,'' {\em EPL}, vol.~135, p.~10003, 2021.

\bibitem{Liu:2022dcn}
W.~Liu, X.~Fang, J.~Jing, and J.~Wang, ``{QNMs of slowly rotating
  Einstein\textendash{}Bumblebee black hole},'' {\em Eur. Phys. J. C}, vol.~83,
  no.~1, p.~83, 2023.

\bibitem{Khodadi:2023yiw}
M.~Khodadi, G.~Lambiase, and L.~Mastrototaro, ``{Spontaneous Lorentz symmetry
  breaking effects on GRBs jets arising from neutrino pair annihilation process
  near a black hole},'' {\em Eur. Phys. J. C}, vol.~83, no.~3, p.~239, 2023.

\bibitem{araujo2025does}
A.~A. Ara{\'u}jo~Filho, ``How does non-metricity affect particle creation and
  evaporation in bumblebee gravity?,'' {\em arXiv e-prints}, pp.~arXiv--2501,
  2025.

\bibitem{AraujoFilho:2024ctw}
A.~A. Ara\'ujo~Filho, ``{Particle creation and evaporation in Kalb-Ramond
  gravity},'' {\em JCAP}, vol.~04, p.~076, 2025.

\bibitem{gravitationaltraces}
A.~A. Ara{\'u}jo~Filho, H.~Hassanabadi, N.~Heidari, J.~Kriz, and S.~Zare,
  ``Gravitational traces of bumblebee gravity in metric--affine formalism,''
  {\em Classical and Quantum Gravity}, vol.~41, no.~5, p.~055003, 2024.

\bibitem{araujo2024gravitational}
A.~A. Ara{\'u}jo~Filho, J.~R. Nascimento, A.~Y. Petrov, P.~J. Porf{\i}rio, {\em
  et~al.}, ``Gravitational lensing by a lorentz-violating black hole,'' {\em
  arXiv preprint arXiv:2404.04176}, 2024.

\bibitem{Gao:2024ejs}
X.-J. Gao, ``{Gravitational lensing and shadow by a Schwarzschild-like black
  hole in metric-affine bumblebee gravity},'' {\em Eur. Phys. J. C}, vol.~84,
  no.~9, p.~973, 2024.

\bibitem{Panotopoulos:2024jtn}
G.~Panotopoulos and A.~{\"O}vg{\"u}n, ``Strange quark stars and condensate dark
  stars in bumblebee gravity,'' {\em Nuclear Physics B}, vol.~1017, p.~116956,
  2025.

\bibitem{Lambiase:2023zeo}
G.~Lambiase, L.~Mastrototaro, R.~C. Pantig, and A.~Ovgun, ``{Probing
  Schwarzschild-like black holes in metric-affine bumblebee gravity with
  accretion disk, deflection angle, greybody bounds, and neutrino
  propagation},'' {\em JCAP}, vol.~12, p.~026, 2023.

\bibitem{neu42}
P.~D. Group, P.~Zyla, R.~Barnett, J.~Beringer, O.~Dahl, D.~Dwyer, D.~Groom,
  C.-J. Lin, K.~Lugovsky, E.~Pianori, {\em et~al.}, ``Review of particle
  physics,'' {\em Progress of Theoretical and Experimental Physics}, vol.~2020,
  no.~8, p.~083C01, 2020.

\bibitem{neu43}
I.~Esteban, M.~C. Gonz{\'a}lez-Garc{\'\i}a, A.~Hernandez-Cabezudo, M.~Maltoni,
  and T.~Schwetz, ``Global analysis of three-flavour neutrino oscillations:
  synergies and tensions in the determination of $\theta$23, $\delta$cp, and
  the mass ordering,'' {\em Journal of High Energy Physics}, vol.~2019, no.~1,
  pp.~1--35, 2019.

\bibitem{neu44}
F.~An, J.~Bai, A.~Balantekin, H.~Band, D.~Beavis, W.~Beriguete, M.~Bishai,
  S.~Blyth, K.~Boddy, R.~Brown, {\em et~al.}, ``Observation of
  electron-antineutrino disappearance at daya bay,'' {\em Physical Review
  Letters}, vol.~108, no.~17, p.~171803, 2012.

\bibitem{neu40}
P.~F. de~Salas, D.~V. Forero, C.~A. Ternes, M.~Tortola, and J.~W.~F. Valle,
  ``{Status of neutrino oscillations 2018: 3$\sigma$ hint for normal mass
  ordering and improved CP sensitivity},'' {\em Phys. Lett. B}, vol.~782,
  pp.~633--640, 2018.

\bibitem{neu39}
F.~Capozzi, G.~L. Fogli, E.~Lisi, A.~Marrone, D.~Montanino, and A.~Palazzo,
  ``{Status of three-neutrino oscillation parameters, circa 2013},'' {\em Phys.
  Rev. D}, vol.~89, p.~093018, 2014.

\bibitem{neu41}
I.~Esteban, M.~C. Gonz{\'a}lez-Garc{\'\i}a, A.~Hernandez-Cabezudo, M.~Maltoni,
  and T.~Schwetz, ``Global analysis of three-flavour neutrino oscillations:
  synergies and tensions in the determination of $\theta$23, $\delta$cp, and
  the mass ordering,'' {\em Journal of High Energy Physics}, vol.~2019, no.~1,
  pp.~1--35, 2019.

\bibitem{neu45}
P.~D. Group {\em et~al.}, ``Review of particle physics,'' {\em Physical Review
  D}, vol.~98, no.~3, p.~030001, 2018.

\bibitem{Shi:2025rfq}
Y.~Shi {\em et~al.}, ``Influence of a kalb-ramond black hole on neutrino
  behavior,'' {\em Journal of High Energy Physics}, vol.~2025, no.~8,
  pp.~1--27, 2025.

\bibitem{Shi:2025plr}
Y.~Shi and A.~Ara{\'u}jo~Filho, ``Effects of bumblebee gravity on neutrino
  motion,'' {\em Journal of Cosmology and Astroparticle Physics}, vol.~2025,
  no.~11, p.~045, 2025.

\bibitem{Alloqulov:2024sns}
M.~Alloqulov, H.~Chakrabarty, D.~Malafarina, B.~Ahmedov, and A.~Abdujabbarov,
  ``{Gravitational lensing of neutrinos in parametrized black hole
  spacetimes},'' {\em JCAP}, vol.~02, p.~070, 2025.

\bibitem{Shi:2024flw}
Y.~Shi and H.~Cheng, ``The neutrino flavor oscillations in the static and
  spherically symmetric black-hole-like wormholes,'' {\em The European Physical
  Journal C}, vol.~85, no.~8, p.~909, 2025.

\bibitem{Chakrabarty:2023kld}
H.~Chakrabarty, A.~Chatrabhuti, D.~Malafarina, B.~Silasan, and T.~Tangphati,
  ``{Effects of gravitational lensing by Kaluza-Klein black holes on neutrino
  oscillations},'' {\em JCAP}, vol.~08, p.~018, 2023.

\bibitem{AraujoFilho:2025rzh}
N.~Heidari, Y.~Shi, and A.~Ara{\'u}jo~Filho, ``Neutrino dynamics in a
  non-commutative spacetime,'' {\em arXiv preprint arXiv:2504.04474}, 2025.

\bibitem{Shi:2023kid}
Y.~Shi and H.~Cheng, ``{The shadow and gamma-ray bursts of a Schwarzschild
  black hole in asymptotic safety},'' {\em Commun. Theor. Phys.}, vol.~77,
  no.~2, p.~025401, 2025.

\bibitem{Shi:2023hbw}
Y.~Shi and H.~Cheng, ``{The gamma-ray burst arising from neutrino pair
  annihilation in the static and spherically symmetric black-hole-like
  wormholes},'' {\em JCAP}, vol.~10, p.~062, 2023.

\bibitem{neu46}
D.~V. Ahluwalia and C.~Burgard, ``Gravitationally induced neutrino-oscillation
  phases,'' {\em General Relativity and Gravitation}, vol.~28, pp.~1161--1170,
  1996.

\bibitem{neu53}
H.~Swami, K.~Lochan, and K.~M. Patel, ``Signature of neutrino mass hierarchy in
  gravitational lensing,'' {\em Physical Review D}, vol.~102, no.~2, p.~024043,
  2020.

\bibitem{neu50}
O.~Luongo and G.~V. Stagno, ``{Neutrino oscillation at the lifshitz point},''
  {\em Mod. Phys. Lett. A}, vol.~26, pp.~1257--1266, 2011.

\bibitem{neu47}
D.~V. Ahluwalia and C.~Burgard, ``About the interpretation of gravitationally
  induced neutrino oscillation phases,'' {\em arXiv preprint gr-qc/9606031},
  1996.

\bibitem{neu49}
T.~Bhattacharya, S.~Habib, and E.~Mottola, ``{Gravitationally induced neutrino
  oscillation phases in static space-times},'' {\em Phys. Rev. D}, vol.~59,
  p.~067301, 1999.

\bibitem{neu48}
Y.~Grossman and H.~J. Lipkin, ``{Flavor oscillations from a spatially localized
  source: A Simple general treatment},'' {\em Phys. Rev. D}, vol.~55,
  pp.~2760--2767, 1997.

\bibitem{neu52}
G.~Koutsoumbas and D.~Metaxas, ``{Neutrino oscillations in gravitational and
  cosmological backgrounds},'' {\em Gen. Rel. Grav.}, vol.~52, no.~10, p.~102,
  2020.

\bibitem{neu51}
A.~Geralico and O.~Luongo, ``{Neutrino oscillations in the field of a rotating
  deformed mass},'' {\em Phys. Lett. A}, vol.~376, pp.~1239--1243, 2012.

\bibitem{neu54}
C.~Y. Cardall and G.~M. Fuller, ``Neutrino oscillations in curved spacetime: A
  heuristic treatment,'' {\em Physical Review D}, vol.~55, no.~12, p.~7960,
  1997.

\bibitem{neu55}
C.~Y. Cardall and G.~M. Fuller, ``Neutrino oscillations in curved spacetime: A
  heuristic treatment,'' {\em Physical Review D}, vol.~55, no.~12, p.~7960,
  1997.

\bibitem{neu58}
M.~Dvornikov, ``{Spin effects in neutrino gravitational scattering},'' {\em
  Phys. Rev. D}, vol.~101, no.~5, p.~056018, 2020.

\bibitem{neu57}
J.~Alexandre and K.~Clough, ``{Black hole interference patterns in flavor
  oscillations},'' {\em Phys. Rev. D}, vol.~98, no.~4, p.~043004, 2018.

\bibitem{neu56}
R.~M. Crocker, C.~Giunti, and D.~J. Mortlock, ``Neutrino interferometry in
  curved spacetime,'' {\em Physical Review D}, vol.~69, no.~6, p.~063008, 2004.

\bibitem{neu59}
J.~Zhang, M.~Liu, Z.~Liu, and S.~Yang, ``{A new touch temperature of the event
  horizon and Rindler horizon in the Kinnersley spacetime},'' {\em Eur. Phys.
  J. C}, vol.~82, no.~1, p.~1, 2022.

\bibitem{Salmonson:1999es}
J.~D. Salmonson and J.~R. Wilson, ``{General relativistic augmentation of
  neutrino pair annihilation energy deposition near neutron stars},'' {\em
  Astrophys. J.}, vol.~517, pp.~859--865, 1999.

\bibitem{AraujoFilho:2024mvz}
A.~Ara{\'u}jo~Filho, N.~Heidari, and A.~{\"O}vg{\"u}n, ``Geodesics, accretion
  disk, gravitational lensing, time delay, and effects on neutrinos induced by
  a non-commutative black hole,'' {\em Journal of Cosmology and Astroparticle
  Physics}, vol.~2025, no.~06, p.~062, 2025.

\bibitem{Lambiase:2020iul}
G.~Lambiase and L.~Mastrototaro, ``{Effects of modified theories of gravity on
  neutrino pair annihilation energy deposition near neutron stars},'' {\em
  Astrophys. J.}, vol.~904, no.~1, p.~19, 2020.

\bibitem{shi2022neutrino}
Y.~Shi and H.~Cheng, ``{The neutrino pair annihilation around a massive source
  with an f(R) global monopole},'' {\em EPL}, vol.~140, no.~4, p.~49001, 2022.

\bibitem{neu18}
Y.~Nambu, S.~Noda, and Y.~Sakai, ``Wave optics in spacetimes with compact
  gravitating object,'' {\em Physical Review D}, vol.~100, no.~6, p.~064037,
  2019.

\bibitem{neu60}
H.~Chakrabarty, D.~Borah, A.~Abdujabbarov, D.~Malafarina, and B.~Ahmedov,
  ``Effects of gravitational lensing on neutrino oscillation in
  $\gamma$-spacetime,'' {\em The European Physical Journal C}, vol.~82, no.~1,
  p.~24, 2022.

\bibitem{neu62}
Z.~Maki, M.~Nakagawa, and S.~Sakata, ``Remarks on the unified model of
  elementary particles,'' {\em Progress of Theoretical Physics}, vol.~28,
  no.~5, pp.~870--880, 1962.

\bibitem{neu61}
B.~Pontecorvo, ``Inverse $ beta $ processes and nonconservation of lepton
  charge,'' {\em Zhur. Eksptl'. i Teoret. Fiz.}, vol.~34, 1958.

\bibitem{neu63}
B.~Pontecorvo, ``Neutrino experiments and the problem of conservation of
  leptonic charge,'' {\em Sov. Phys. JETP}, vol.~26, no.~984-988, p.~165, 1968.

\bibitem{swami2020signature}
H.~Swami, K.~Lochan, and K.~M. Patel, ``Signature of neutrino mass hierarchy in
  gravitational lensing,'' {\em Physical Review D}, vol.~102, no.~2, p.~024043,
  2020.

\bibitem{neu64}
L.~Stodolsky, ``Matter and light wave interferometry in gravitational fields,''
  {\em General Relativity and Gravitation}, vol.~11, pp.~391--405, 1979.

\bibitem{neu65}
H.~Swami, K.~Lochan, and K.~M. Patel, ``Aspects of gravitational decoherence in
  neutrino lensing,'' {\em Physical Review D}, vol.~104, no.~9, p.~095007,
  2021.

\bibitem{Barros:2025din}
P.~H.~M. Barros, F.~C.~E. Lima, C.~A.~S. Almeida, and H.~A.~S. Costa,
  ``{Mitigating the information degradation in a massive Unruh-DeWitt
  theory},'' {\em JHEP}, vol.~04, p.~165, 2025.

\bibitem{Barros:2024wdv}
P.~H.~M. Barros and H.~A.~S. Costa, ``{Detecting gravitational waves via
  coherence degradation induced by the Unruh effect},'' {\em Eur. Phys. J. C},
  vol.~84, no.~12, p.~1261, 2024.

\end{thebibliography}

\end{document}